\documentclass[aps,prl,twocolumn,showpacs,10pt]{revtex4-1}
\usepackage{graphicx}
\usepackage{amssymb}
\usepackage{amsmath,color}
\usepackage{bbold}
\usepackage{comment}
\pdfoutput=1

\begin{document}

\def\pc{\frac{2\pi}{\Phi_0}}

\def\e{\varepsilon}
\def\f{\varphi}
\def\p{\partial}
\def\ba{\mathbf{a}}
\def\bA{\mathbf{A}}
\def\bb{\mathbf{b}}
\def\bB{\mathbf{B}}
\def\bD{\mathbf{D}}
\def\bd{\mathbf{d}}
\def\be{\mathbf{e}}
\def\bE{\mathbf{E}}
\def\bH{\mathbf{H}}
\def\bj{\mathbf{j}}
\def\bk{\mathbf{k}}
\def\bK{\mathbf{K}}
\def\bM{\mathbf{M}}
\def\bm{\mathbf{m}}
\def\bn{\mathbf{n}}
\def\bq{\mathbf{q}}
\def\bp{\mathbf{p}}
\def\bP{\mathbf{P}}
\def\br{\mathbf{r}}
\def\bR{\mathbf{R}}
\def\bS{\mathbf{S}}
\def\bu{\mathbf{u}}
\def\bv{\mathbf{v}}
\def\bV{\mathbf{V}}
\def\bw{\mathbf{w}}
\def\bx{\mathbf{x}}
\def\by{\mathbf{y}}
\def\bz{\mathbf{z}}
\def\bG{\mathbf{G}}
\def\bW{\mathbf{W}}
\def\Bn{\boldsymbol{\nabla}}
\def\Bo{\boldsymbol{\omega}}
\def\Br{\boldsymbol{\rho}}
\def\Bs{\boldsymbol{\hat{\sigma}}}
\def\bh{{\beta\hbar}}
\def\mA{\mathcal{A}}
\def\mB{\mathcal{B}}
\def\mD{\mathcal{D}}
\def\mF{\mathcal{F}}
\def\mG{\mathcal{G}}
\def\mH{\mathcal{H}}
\def\mI{\mathcal{I}}
\def\mL{\mathcal{L}}
\def\mO{\mathcal{O}}
\def\mP{\mathcal{P}}
\def\mT{\mathcal{T}}
\def\mZ{\mathcal{Z}}
\def\fr{\mathfrak{r}}
\def\ft{\mathfrak{t}}
\newcommand{\rf}[1]{(\ref{#1})}
\newcommand{\al}[1]{\begin{aligned}#1\end{aligned}}
\newcommand{\ar}[2]{\begin{array}{#1}#2\end{array}}
\newcommand{\eq}[1]{\begin{equation}#1\end{equation}}
\newcommand{\bra}[1]{\langle{#1}|}
\newcommand{\ket}[1]{|{#1}\rangle}
\newcommand{\av}[1]{\langle{#1}\rangle}
\newcommand{\AV}[1]{\left\langle{#1}\right\rangle}
\newcommand{\aav}[1]{\langle\langle{#1}\rangle\rangle}
\newcommand{\braket}[2]{\langle{#1}|{#2}\rangle}
\newcommand{\ff}[4]{\parbox{#1mm}{\begin{center}\begin{fmfgraph*}(#2,#3)#4\end{fmfgraph*}\end{center}}}

\title{Universal Quantum Computation with Hybrid Spin-Majorana Qubits}

\author{Silas Hoffman$^1$}
\author{Constantin Schrade$^1$}
\author{Jelena Klinovaja$^1$}
\author{Daniel Loss$^1$}
\affiliation{$^1$Department of Physics, University of Basel, Klingelbergstrasse 82, CH-4056 Basel, Switzerland}


\begin{abstract}
We theoretically propose a set of universal  quantum gates acting on a hybrid qubit formed by coupling a quantum dot spin qubit and Majorana fermion qubit. First, we consider a quantum dot tunnel-coupled to two topological superconductors. The effective spin-Majorana exchange facilitates a hybrid CNOT  gate for which either qubit can be the control or target.  The second setup is a modular scalable network of topological superconductors and quantum dots. As a result of the exchange interaction between adjacent spin qubits, a CNOT gate is implemented that acts on neighboring Majorana qubits, and eliminates the necessity of inter-qubit braiding. In both setups the spin-Majorana exchange interaction allows for a phase gate, acting on either the spin or the Majorana qubit, and for a SWAP or hybrid SWAP gate which is sufficient for universal quantum computation without projective measurements.
\end{abstract}
 
\maketitle
\textit{Introduction.} Quantum dots are promising, scalable, settings to store and  manipulate quantum information using  spin states  \cite{lossPRA98,hansonRMP07}. However, the quantum data stored is susceptible to decoherence by the environment wherein quantum information is lost \cite{kloeffelARCMP13}.

An alternative proposal to such traditional quantum bits are topological quantum computers \cite{nayakRMP08} which make use of degenerate ground states of topological matter, whose edge states obey non-Abelian statistics upon exchange \cite{goldinPRL85}, to encode qubits. The information stored in these nonlocal degrees of freedom are tolerant to local system noise and can be manipulated by braiding \cite{kitaevAoP03,freedmanCiMP02, bravyiAoP02,freedmanBAMS03}.
 There are several proposed realizations of such topological qubits~\cite{nayakRMP08}, the most successful one to date being
those composed of Majorana fermions (MFs) due to their immediate experimental accessibility \cite{mourikSCI12,dasNATP12,rokhinsonNATP12,dengNANOL12,finckPRL13,churchillPRB13,nadj-pergeSCI14, pawlakCM15}. Several theoretical setups to realize MFs have been proposed:  semiconducting-superconducting nanowires \cite{lutchynPRL10, oregPRL10}, topological insulators \cite{fuPRL08},
topological superconductors (TSCs)
~\cite{volovikJETP99}, and magnetic adatoms on top of $s$-wave superconductors \cite{klinovajaPRL13,vazifehPRL13,brauneckerPRL13,nadj-pergePRB13,pientkaPRB13}. However,  MFs do not generate a universal set of topological gate operations necessary for quantum computation \cite{bravyiPRA06}.

\begin{figure}
\includegraphics[width=1\linewidth]{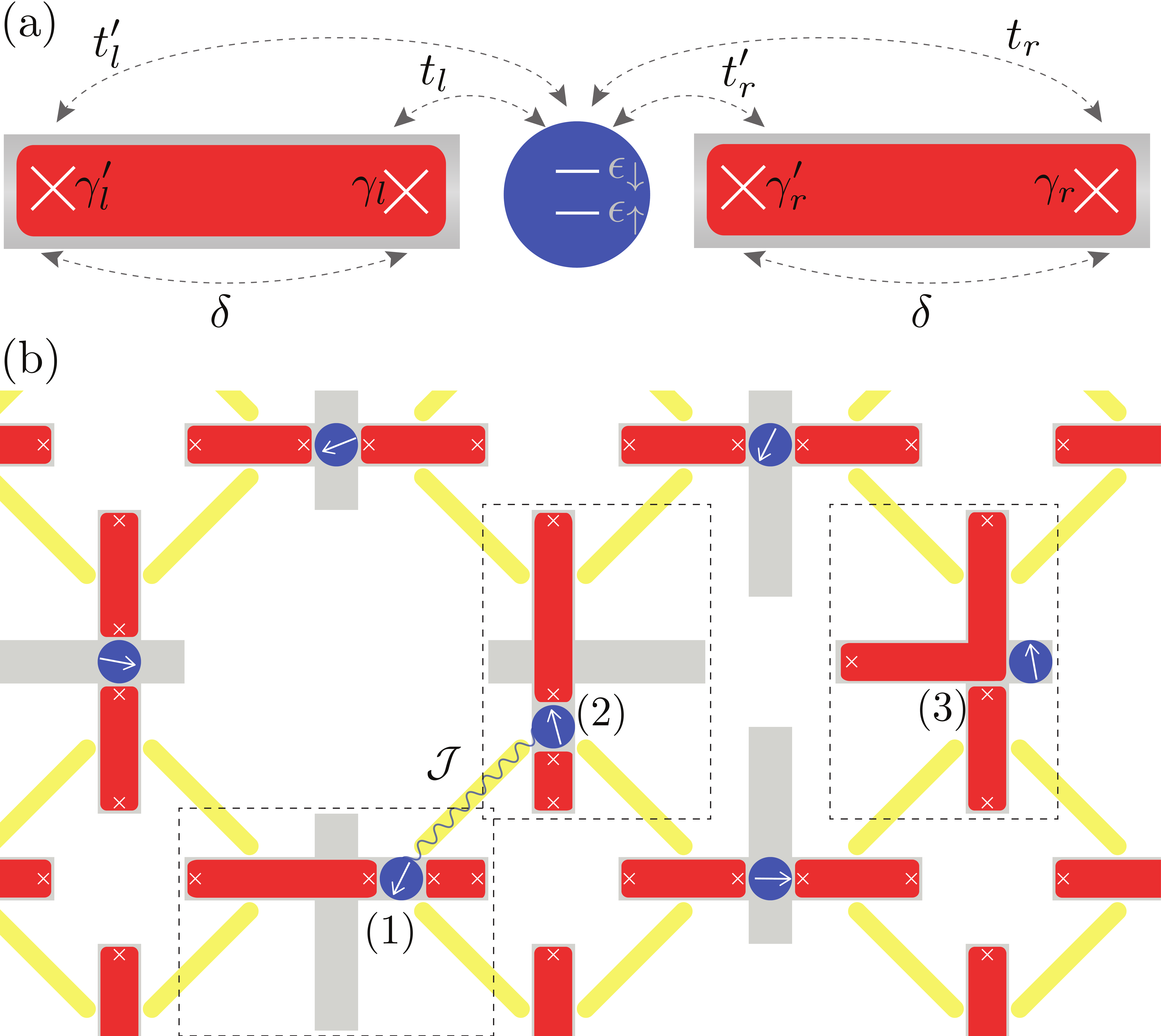}
\caption{(a) Setup of two TSCs (red bars) furnishing two MFs (crosses) on the left TSC, $\gamma'_{l}$ and $\gamma_{l}$, and two MFs on the right TSC, $\gamma'_{r}$ and $\gamma_{r}$; the MFs on each TSC can overlap causing a splitting  $\delta$. Between the two TSCs is a quantum dot (blue disc) with two single electron levels of up, $\epsilon_\uparrow$, and down, $\epsilon_\downarrow$, spin. The MFs are coupled to the dot through the tunneling elements $t_{\nu}$ and $t'_{\nu}$   where $\nu$ labels the right ($r$) and left ($l$) TSCs. (b) MaSH network of TSCs where a grid of hybrid qubits (red and grey crosses) are long-distance coupled by tunably connecting the spin-1/2 quantum dots, with strength $\mathcal{J}$, via floating gates, \textit{e.g.} hybrid qubit (1) is coupled to hybrid qubit (2). Braiding of MFs utilizes the T-junctions of the TSCs on each hybrid qubit; for instance hybrid qubit (3).
}
\label{setup}
\end{figure}

The additional non-topological gates needed to achieve universality with MF qubits can be implemented by fusing anyons \cite{bravyiPRA06}, using magnetic flux \cite{hyartPRB13}, or quantum information transfer with spins in quantum dots \cite{leijnsePRL11}. The principle drawback of these schemes is twofold: (1) after preparing the system state, a projective measurement must be made, which should be perfect~\cite{bravyiPRA06} and which is typically time intensive \cite{hansonRMP07}; (2) braiding between two topological qubits is required to perform universal quantum computation, which necessitates a long distance topologically nontrivial interaction between them. In this Letter, using a hybrid qubit composed of a coupled spin and MF qubit [Fig.~{\ref{setup}(a)], we can coherently transfer information between the qubit components, thereby  keeping the gate operation time on MF qubits potentially as short as possible. Furthermore, when the spins on two such hybrid qubits are allowed to interact, universal quantum computation can be achieved by applying gate operations directly to MF qubits using fixed spin qubits as a control for the interaction, thus eliminating the need for large coherent networks. Making use of such a coupling, we propose a scalable modular network of Majorana and spin hybrid (MaSH) qubits [Fig.~{\ref{setup}(b)].

In the following, we derive the effective coupling between the spin and MF qubits which is used to perform a phase gate on the MF qubit and a SWAP gate between the spin and MF qubits. Extending the system to a network of MaSH qubits, long-distance coupled by the spins, we demonstrate the necessary operations to obtain universal quantum computing. Because MFs can be realized in many different setups, we have considered a rather general coupling between spin and MF qubits which provides a proof of principle for a wide class of physical systems.

\textit{Setup.} We consider a single level quantum dot placed between two TSCs [Fig.~\ref{setup}(a)], which can be realized as any of the previously mentioned setups.
The chemical potential and Coloumb repulsion, $U$, on the dot are assumed to be tuned to favor single occupancy (or more generally a spin-1/2 groundstate). The two opposite spin levels of the dot  $\epsilon_{\uparrow/\downarrow}$ are non-degenerate  in the presence of a magnetic field. The Hamiltonian of the quantum dot is $H_D=\sum_{\sigma=\uparrow,\downarrow}(\epsilon_\sigma d_\sigma^\dagger d_\sigma+Un_\sigma n_{\bar\sigma}/2)$, where $d_\sigma^\dagger$ ($d_\sigma$) creates (annihilates) an electron with spin $\sigma$ and $n_\sigma=d^\dagger_\sigma d_\sigma$. The right ($r$) and left ($l$) TSCs, modeled as a Kitaev chain~\cite{kitaevPU01}, are tuned to the topological regime, furnishing MFs at opposite  ends. As the separation between MFs can be comparable with the MF localization length, we include a phenomenological splitting of $\delta$ between MFs in the same TSC but neglect splitting between MFs on opposite TSCs \cite{rainisPRB13,zyuzinPRL13}. Neglecting also quasiparticle excitations \cite{rainisPRB12,pedrocchiPRL15,hutterCM15}
, we consider the MF states on the TSC, which is a good approximation when the tunneling is much smaller than the superconducting gap; the Hamiltonian of the TSC is $H_M=\sum_{\nu=r,l}i\delta\gamma_{\nu}'\gamma_{\nu}$, where $\gamma_{\nu}'$ ($\gamma_{\nu}$) is the MF at the left (right) end of the $\nu$th TSC and we have set the chemical potential of the superconductors to zero.

The overlap of the electron wavefunctions on the dot and MF wavefunctions in the TSC is described by the tunneling Hamiltonian \cite{kitaevPU01,tewariPRL08}, $H_T=\sum_{\sigma,\nu}d_\sigma^\dagger(i t'_{\nu}\gamma'_{\nu}+ t_{\nu}\gamma_{\nu})+\textrm{H.c.}$, 
where $t'_{\nu}$ ($t_{\nu}$) is the matrix element for an electron on the dot tunneling into the left (right) MF in the $\nu$th TSC . We assume our Kitaev chains to have a single spin species oriented perpendicular to the axis of quantization on the dot and the tunneling elements to be spin independent. A spin dependent tunneling, or equivalently choosing a different axis of spin polarization on the TSC, changes the direction of the effective magnetic field on the dot \cite{braunPRB04}, which should not qualitatively affect our results.

Each pair of MFs in the TSCs are conveniently described as a single Dirac fermion $f_\nu=(\gamma'_{\nu}+i\gamma_{\nu})/2$; the MF and tunneling Hamiltonians are rewritten as $H_M=\sum_\nu\delta(2f_\nu^\dagger f_\nu -1)$ and $H_T=\sum_{\sigma,\nu}i t_{\nu -}^*f^\dagger_\nu d_\sigma - i t_{\nu +}^*f_{\nu} d_\sigma+\textrm{H.c.}$, respectively, where $t_{\nu \pm}=t_{\nu}\pm t'_{\nu}$. The value of $f_\nu^\dagger f_\nu=0,1$ determines the parity of the $\nu$th TSC,  which can be even or odd, respectively. As the dot is always singly-occupied, the total parity of the MF qubit, defined as the sum of the parities of the TSCs modulo two, is fixed to be in an even- or odd-parity subspace~\cite{SM} of the full Hilbert space. The terms proportional to $t_{\nu +}$ ($t^*_{\nu +}
$) correspond to removing (adding) a Cooper pair from the condensate and adding (removing) one electron to the dot and one to the $\nu$th TSC; the terms proportional to $t_{\nu -}$ or $t^*_{\nu -}
$ correspond to the transfer of electrons between the dot and the $\nu$th TSC \cite{tewariPRL08}. The full model Hamiltonian of our hybrid qubit system is  $H=H_D+H_M+H_T$. 

\begin{figure}
\includegraphics[width=1\linewidth]{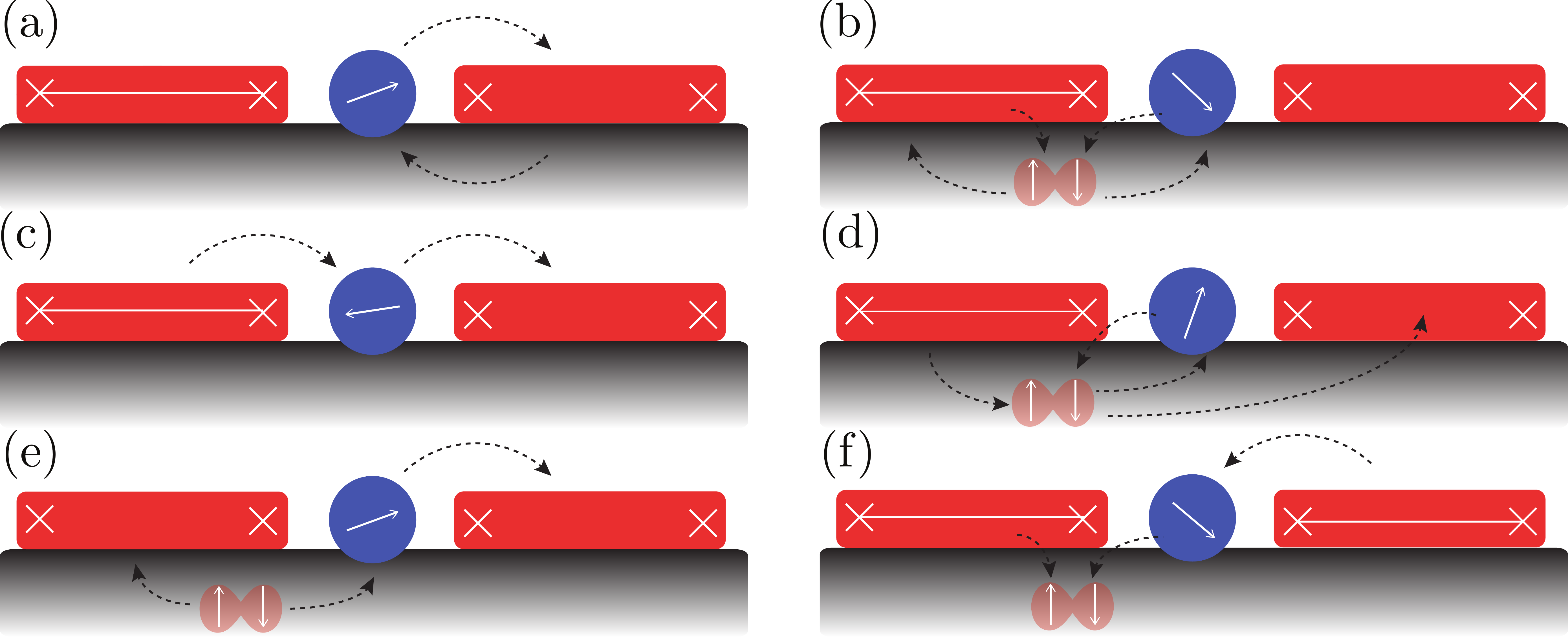}
\caption{Some of the processes that result from the coupling between spin and MF qubit dictated by $\mathcal H_T$. 
The straight line (white) connecting the MFs (crosses) indicates odd parity, {\it i.e.}, $f^\dagger_\nu f_\nu =  1$.
(a-b) The virtual processes described by $\mathcal H_s$; the remaining undepicted processes are similar but take place on the one, three, and four total electron state. (c-d) The transfer of an electron from one TSC to the other due to $\mathcal{H}_o$. (e-f) The processes  determined by $\mathcal H_e$ that map the system between the two states in the even parity sector of the MF qubit. The other processes resulting from $\mathcal H_o$ and $\mathcal H_e$ are obtained by exchanging the right and left TSCs (or initial and final states) in panels (c-f).}
\label{virt}
\end{figure}

\textit{Effective Hamiltonian.} If the coupling between the dot and the TSCs is weak compared to the difference in energies of the dot electrons and MFs, we obtain an effective Hamiltonian $\mathcal H_T=\mathcal H_s+\mathcal H_o+\mathcal H_e$ by applying a Schrieffer-Wolff transformation~\cite{schriefferPR66,leePRB13,vernekPRB14} to $H$ (see  also SM~\cite{SM}), 
\begin{align}
&\mathcal H_s=\sum_{\sigma,\nu}\left(\frac{| t_{\nu -}|^2}{\epsilon_\sigma-2\delta}f_\nu f_\nu^\dagger+\frac{| t_{\nu +}|^2}{\epsilon_\sigma+2\delta}f_\nu^\dagger f_\nu\right)\mathcal B_\sigma \label{Hsw} \\
&\mathcal H_o=\sum_{\sigma,\nu}\left(\frac{ t^*_{\bar\nu -} t_{\nu -}}{\epsilon_\sigma-2\delta} f_\nu f_{\bar\nu}^\dagger+\frac{ t^*_{\bar\nu +} t_{\nu +}}{\epsilon_{\sigma}+2\delta} f_\nu^\dagger f_{\bar\nu}\right)\mathcal B_\sigma\nonumber\\
&\mathcal H_e=-\sum_{\sigma,\nu} t_{\bar\nu -}^* t_{\nu +}\left(\frac{\mathcal A_\sigma}{\epsilon_\sigma-2\delta}+\frac{\mathcal  A^{\dagger}_\sigma}{\epsilon_{\sigma}+2\delta}\right)  f_\nu^\dagger f_{\bar\nu}^\dagger +\textrm{H.c.}\nonumber
\end{align}
We have taken $U$ the largest energy scale, {\it i.e.},
$U\rightarrow\infty$, and defined the operators $\mathcal  A_\sigma=n_\sigma + d^\dagger_{\bar\sigma}d_\sigma $ and $\mathcal B_\sigma = \mathcal  A_\sigma + \mathcal  A^{\dagger}_\sigma $. Here, $\mathcal H_s$ results from  hopping between the dot and a single TSC. The term proportional to $|t_{\nu -}|^2$ corresponds to the process of the electron on the dot hopping to the $\nu$th TSC then back to the dot [Fig.~\ref{virt}(a)], while the term proportional to $|t_{\nu +}|^2$ corresponds to the process of the electron on the dot combining with the electron on the $\nu$th TSC  into a Cooper pair, and breaking a Cooper pair adding one electron to the dot and one to the same TSC [Fig.~\ref{virt}(b)]; both processes can happen in either parity subspace. The Hamiltonian $\mathcal H_o$ ($\mathcal H_e$) results from hopping between the dot and both TSCs, 
which couple states in the odd (even) parity subspace exclusively. The term proportional to $t_{\bar\nu -}^*t_{\nu -}$ corresponds to transferring an electron from the dot to the even parity TSC then from the odd parity TSC to the quantum dot [Fig.~\ref{virt}(c)].   The condensing of the electron on the dot and with the electron from the odd parity TSC into a Cooper pair and then breaking apart a Cooper pair, putting one electron on the opposite TSC and the other electron on the dot  [Fig.~\ref{virt}(d)], is described by $t_{\bar\nu +}^*t_{\nu +}$. The term proportional to $t_{\bar\nu -}^*t_{\nu +}$ acts on the  zero total  electron state by transferring the dot electron to the $\nu$th TSC  then taking two electrons from the condensate, filling the state in the latter TSC and transferring the other onto the dot [Fig.~\ref{virt}(e)]. The latter term, $t_{\bar\nu +}^*t_{\nu -}$, acting on the three total  electron state, condenses the dot electron with one of the TSC electrons while the other TSC electron tunnels onto the dot [Fig.~\ref{virt}(f)].  

In order to create a MF qubit, one must have a superposition of same parity states. In the two TSC system, we restrict to the even total parity or odd MF qubit parity subspace, \textit{i.e.}, one electron on the dot and one electron on either the right ($|r\rangle=f^\dagger_r|0\rangle$) or left ($|l\rangle=f^\dagger_l|0\rangle$) TSC with $|0\rangle$ being the vacuum. In first quantized notation, the effective  Hamiltonian is $\mathcal H_T=\sum_{\kappa,\lambda=0,\ldots,4} J_{\kappa\lambda} \sigma_\kappa \eta_\lambda$, where
$\sigma_\kappa$ ($\eta_\lambda$) act on the spin of the dot (odd parity sector of TCSs defined such that $\eta_3|r\rangle=+|r\rangle$ and $\eta_3|l\rangle=-|l\rangle$).  For $\kappa~(\lambda)\in \{1,2,3\}$ these are  the standard Pauli matrices, while  $\sigma_0~(\eta_0)$ is the identity matrix. The anisotropic exchange constant $J_{\kappa\lambda}$ is a function of $\delta$, $\epsilon_\sigma$, and $t_{\nu\pm}$ \cite{SM}. 

\begin{figure}
\includegraphics[width=.85\linewidth]{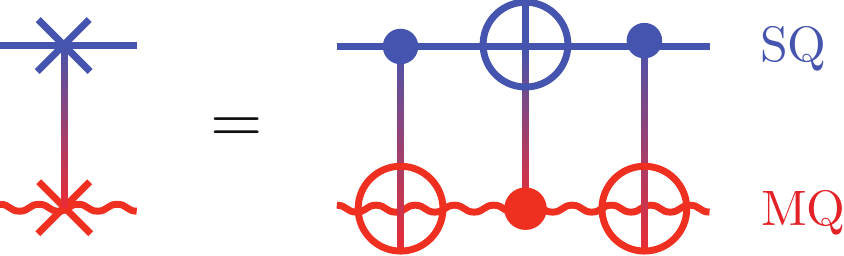}
\caption{Schematic of the hybrid swap (hSWAP) gate obtained as follows: apply the hCNOT gate using, say, the spin qubit (SQ) as the control and the MF qubit (MQ) as the target qubit, apply the hCNOT gate reversing the roles of the qubits, apply the hCNOT gate with the control and target qubits as in the first operation. Starting with the initial state $|x,y\rangle$ such that $x,~y\in\{0,\ 1\}$, where we identify $0$ ($1$) with the $\left|\downarrow\right\rangle$ ($\left |\uparrow\right\rangle$) and $|l\rangle$ ($|r\rangle$) state of the spin and MF qubit, respectively, applying the pictured gate sequence one obtains $|x,y\rangle\rightarrow(-1)^x|x,y\oplus x\rangle\rightarrow(-1)^y|y,y\oplus x\rangle\rightarrow|y,x\rangle$; this results in coherent swap of states between the spin and MF qubit. }
\label{gate}
\end{figure}

\textit{Quantum Gates.} In general,  when the interaction between qubits is entangling, \textit{i.e.,} $J_{\kappa\lambda}\neq0$ for $\kappa\,,\lambda\neq0$, a SWAP gate between the qubits can be implemented. However, a simple setup that yields a so-called hybrid SWAP (hSWAP) gate (Fig.~\ref{gate}) consists of two semi-infinite TSCs with no magnetic field on the dot. The first condition implies that the outer MF wave functions do not overlap with that of the inner MFs ($\delta=0$) or the quantum dot ($t_{l}'=t_{r}=0$), while the second implies the spin states on the dot are degenerate, $\epsilon_\uparrow=\epsilon_\downarrow=\epsilon_0$, for which $\mathcal H_T$ becomes $(\mathbb 1+\sigma_1)\left[(|t_r'|^2+|t_l|^2)+2\textrm{Re} (t_r' t_l^*) \eta_1\right]/\epsilon_0$.
When $t_l=t_r'=t$, $\mathcal H_T$ further reduces to \cite{lossPRA98,nielsenBK10,SM}
\eq{
\mathcal H_{\textrm{hCP}}=2 |t|^2(\mathbb 1+\sigma_1)(\mathbb1+\eta_1)/\epsilon_0\,,
\label{HCP}
}
which can be used to perform a hybrid controlled phase (hCP) gate. Although in the following we focus on the manipulation of the MF qubit using the spin qubit, owing to the symmetry of the Hamiltonian between spin and MF operations, \textit{i.e.,} under the exchange $\sigma_1\leftrightarrow\eta_1$, one could equally use the MF qubit to manipulate the spin qubit.

After a single-qubit unitary rotation by a Hadamard gate, which can be implemented by applying a magnetic field to the spin qubit and by braiding \cite{bravyiPRA06}  MFs  [Fig.~\ref{MF_gates}(a)], $\mathcal H_{\textrm{hCP}}$ transforms into $\mathcal H^{ij}_{\textrm{hCNOT}} =2 |t|^2(\mathbb 1+\sigma_i)(\mathbb 1+\eta_j)/\epsilon_0$, where $(i,j)=(1,3)$ or $(3,1)$.
Pulsing the coupling $t$ between the dot and TSCs for the duration $\tau$ so that $\int^\tau \mathcal H^{ij}_{\textrm{hCNOT}}=\pi(\mathbb1+\sigma_i)(\mathbb 1+\eta_j)/4$, one obtains the hybrid CNOT (hCNOT) gate $U^{ij}_\textrm{hCNOT}=(\mathbb1-\sigma_i-\eta_j-\sigma_i \eta_j)/2$~\cite{SM} from which an hSWAP gate can be coded as $U_{\textrm{hSWAP}}=U^{31}_\textrm{hCNOT}U^{13}_\textrm{hCNOT}U^{31}_\textrm{hCNOT}$. Applying the hSWAP gate to the two qubits exchanges the relative weights of the up and down spin states of the spin qubit with the right and left parity states of the MF qubit (Fig.~\ref{gate}), respectively.
To implement a $\pi/8$ gate, one may hSWAP the quantum state of the MF qubit onto the spin qubit, perform a $\pi/8$ gate on the spin qubit, and hSWAP the states back; this requires no preparation or projective measurement. Alternatively, one can fix the spin qubit by a magnetic field along the $z$ axis and pulse $H_{\textrm{CNOT}}^{13}$. This generates a phase gate for any value of phase according to the duration of the pulse [Fig.~\ref{MF_gates}(b)]. These three gates are sufficient for universal quantum computation of the hybrid qubit.

\textit{MaSH Network.} We consider a network of MaSH qubits formed by crossing one TSC in the topological phase with one in the trivial phase and defining the spin-1/2 quantum dots at their intersection [Fig.~\ref{setup}(b)]. The MaSH qubit elements are connected via floating gates \cite{trifunovicPRX12} whose ends are placed off center from quantum dots. One can perform braiding of MFs as usual~\cite{aliceaNATP11} by moving the quantum dot to an unused topologically trivial arm of the hybrid qubit so it does not participate in the operation. Because the coupling of quantum dots through floating gates is very sensitive to the relative position of the two \cite{trifunovicPRX12}, the hybrid qubits are engaged when the spin qubit components are near the respective edges of the connective floating gate. 
This induces an isotropic interaction given by $\mathcal J{\vec \sigma}^{(i)}\cdot {\vec \sigma}^{(j)}$, where $(i,j)$ refers to two neighboring hybrid qubits, say $i=1$ and $j=2$. If $\mathcal J\gg|t|$, there is an effective interaction between the MF qubits modulated by the relative direction of the spin qubits,
\eq{
\mathcal H^{(12)}_{MQ}=\frac{|t|^4}{\epsilon_0^2\mathcal J}\left[\sigma^{(1)}_2\sigma^{(2)}_2+\sigma^{(1)}_3\sigma^{(2)}_3\right]\left[1-\eta_1^{(1)}\eta_1^{(2)}\right]\,.
}
Fixing the direction of the spin qubits along the $z$ axis and applying $\mathcal H^{(12)}$ for a specified time \cite{SM}, one obtains the gate $U_{MQ}^{(12)}=\exp[i\pi(1-\eta_1^{(1)}\eta_1^{(2)})/4]$, which directly couples the two MF qubits.  A CNOT gate [Fig.~\ref{MF_gates}(c)], using MF qubit $(1)$ as the target and qubit $(2)$ as the control, can be implemented using the sequence 
\eq{
U^{(12)}_{\textrm{CNOT}}=H^{(2)}U^{(12)}_{MQ}H^{(1)}H^{(2)}R^{(1)}R^{(2)}H^{(1)},
}
where $H^{(i)}$ and $R^{(i)}$ are the Hadamard and $(-\pi/4)$-phase gates, respectively,
acting on the  $i$th MF qubit \cite{SM}. Therefore, using this CNOT gate, the Hadamard  and $\pi/8$ gate in the MaSH setup,  one can implement the necessary gates to realize universal quantum computation by fixing the spin qubits as a control and storing all quantum information in the MF qubits. As noted before, owing to the symmetry of the setup, the role of the spin and the MF qubits can be interchanged and the MF qubits can be used as control qubits.
One may also use the spin qubit to read out parity of the MF qubit by applying the hSWAP gate and measuring the spin on the dot. A simpler alternative to MF parity readout is given in the Supplementary Material~\cite{SM}. Finally, this network can serve as a platform for the surface code with the well-known error threshold of 1.1\% \cite{raussendorfPRL07,wangPRA11,trifunovicPRX12}.
\begin{figure}
\includegraphics[width=1\linewidth]{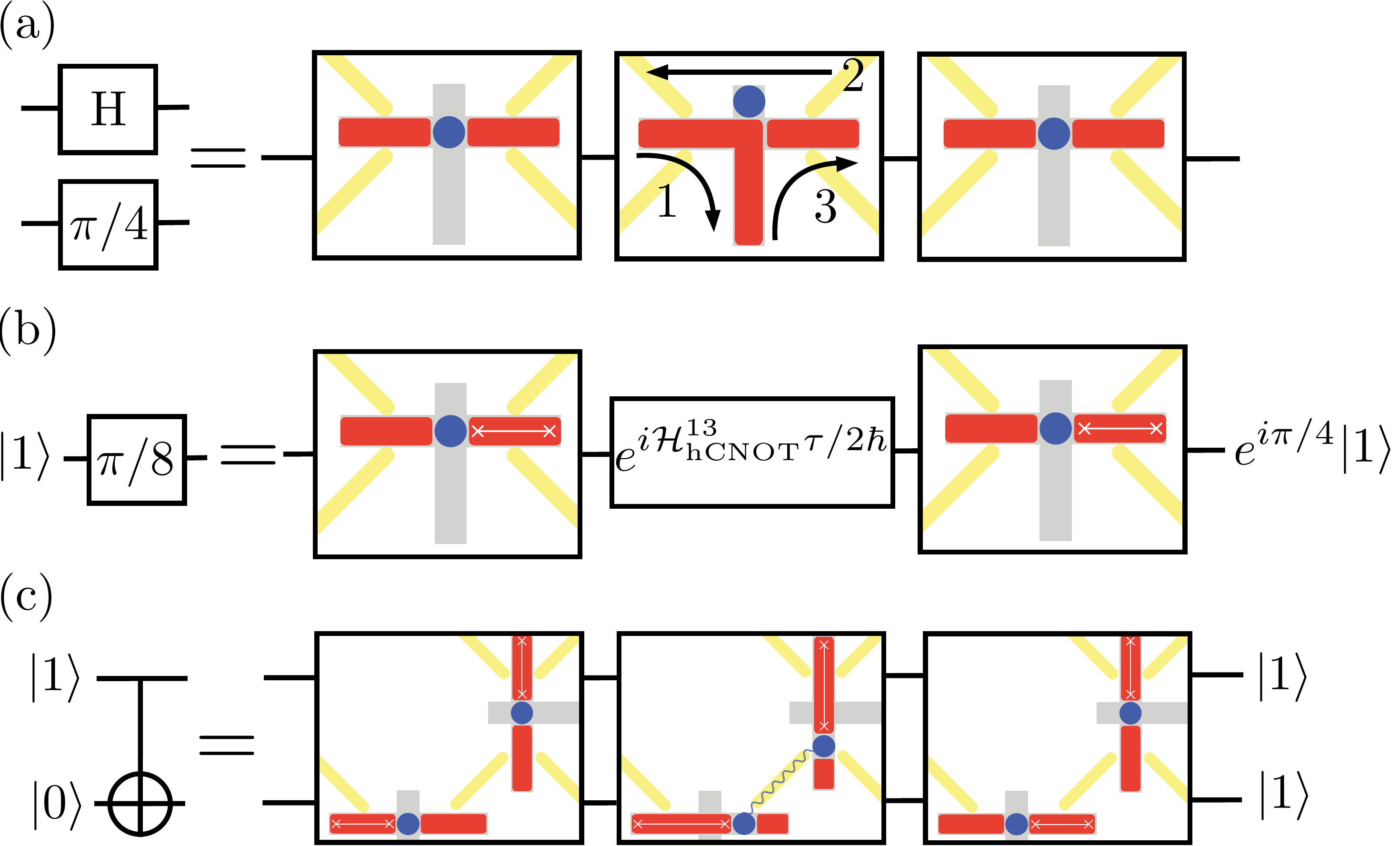}
\caption{Implementation of the necessary gates for universal quantum computation: (a) Hadamard 
and $\pi/4$ phase 
gate as a result of braiding, (b) $\pi/8$ phase gate obtained by coupling the MF qubit and the fixed spin qubit, and (c) CNOT gate obtained through an effective coupling of two MF qubits facilitated by a long-range interaction between the corresponding spin qubits. }
\label{MF_gates}
\end{figure}

\textit{Outlook.} Although there are several systems in which our setup could be implemented, perhaps the most natural scenario is in nanowires because (1) signatures of MFs in nanowires with proximity-induced superconductivity were identified  experimentally \cite{mourikSCI12,leePRL12,dasNATP12,rokhinsonNATP12,dengNANOL12,finckPRL13,churchillPRB13}; (2) single electron quantum dots and electrical implementation of single qubit quantum gates were realized in semiconducting nanowires \cite{fasthPRL07,nadj-pergeNAT10,nadj-pergePRL12} 
%and in nanowires 
also on top of superconductors \cite{defranceschiNATN10,leePRL12,szombatiCM15}.

For a single hybrid qubit setup, we envision one nanowire on top of a conventional $s$-wave superconductor in which one electrically tunes the left and right ends of the wire into the topological regime while a quantum dot  is electrically defined between them. The length of the topological section in the wires can be changed, thereby independently controlling the overlap between the MFs ($\delta$). Similarly, one may set the size of the quantum dot so that the Coulomb repulsion is large as well as applying a gate voltage to ensure the dot is 
in a spin-1/2 groundstate
and fix the dot energy level ($\epsilon_\sigma$) relative to the chemical potential of the wires. One can likewise control the tunneling between quantum dot and wire ($t_{\nu}$ and $t_{\nu}'$) by either adjusting the distance between the two or tuning the barrier height that separates them. To assemble a MaSH network, one composes individual hybrid qubits from two crossed nanowires then connects them with floating gates. Voltage controls, in addition to the previously mentioned tunneling elements, braiding operations and the position of the quantum dot and thus the effective coupling between spin qubits ($\mathcal J$). 

\textit{Conclusions.} By coupling spin and Majorana qubits, we have constructed the necessary gates for universal quantum computation of spin-Majorana hybrid qubits. Forming a MaSH network, a universal set of gates can be implemented directly on the MF qubits while using the spin qubits only as a control. Thanks to the modular nature of this setup and the construction of the CNOT gate, it is unnecessary to engineer a large scale coherent network of TSCs. The necessary experimental techniques to realize a single spin-MF hybrid qubit or a network of such qubits are available. Our results demonstrate that one can harness universal  quantum computation from both single and multiple element spin-MF hybrid qubit systems. 

\textit{Acknowledgements.} This work is supported by Swiss NSF and NCCR QSIT. We would like to acknowledge interesting discussions with J. R. Wootton.

\begin{widetext}

\newpage

\onecolumngrid

\bigskip 

\begin{center}
\large{\bf Supplemental Material to `Universal Quantum Computation with Hybrid Spin-Majorana Qubits' \\}
\end{center}
\begin{center}
Silas Hoffman$^1$, Jelena Klinovaja$^1$, Constantin Schrade$^1$, and Daniel Loss$^{1}$\\
{\it $^1$Department of Physics, University of Basel, Klingelbergstrasse 82, CH-4056 Basel, Switzerland}
\end{center}

\section{Effective Hamiltonian}
In this section we perform a Schrieffer-Wolff transformation \cite{schriefferPR66} on the tunneling Hamiltonian beginning with a Hamiltonian that couples two Kitaev chains to a quantum dot,
\begin{align}
&H=H_M+H_D+H_T\,,\nonumber\\
&H_M=i\sum_\nu\delta_\nu\gamma_{\nu}'\gamma_{\nu}\,,\nonumber\\
&H_D=\sum_\sigma\epsilon_\sigma d_\sigma^\dagger d_\sigma+Un_\sigma n_{\bar\sigma}/2\,,\nonumber\\
&H_T=\sum_{\sigma,\nu}d_\sigma^\dagger(it_{\nu}'\gamma_{\nu}'+t_{\nu}\gamma_{\nu})+(t^*_{\nu}\gamma_{\nu}-it'^*_{\nu}\gamma_{\nu}')d_\sigma \,,
\end{align}
where $\nu$ labels the left ($l$) and right ($r$) chains. We rewrite the Majorana fermions as $f_\nu=(\gamma_{\nu}'+i\gamma_{\nu})/2$ so that $f^\dagger_\nu f_\nu=(1+i\gamma'_{\nu}\gamma_{\nu})/2$ and $i\delta_\nu\gamma_{\nu}'\gamma_{R\nu}=\delta_\nu(2f_\nu^\dagger f_\nu -1)$. The logical values of the qubit are written in terms of the parity of the left and right TSCs. See main text and Fig.~\ref{parity}.
\begin{figure}
\includegraphics[width=1\linewidth]{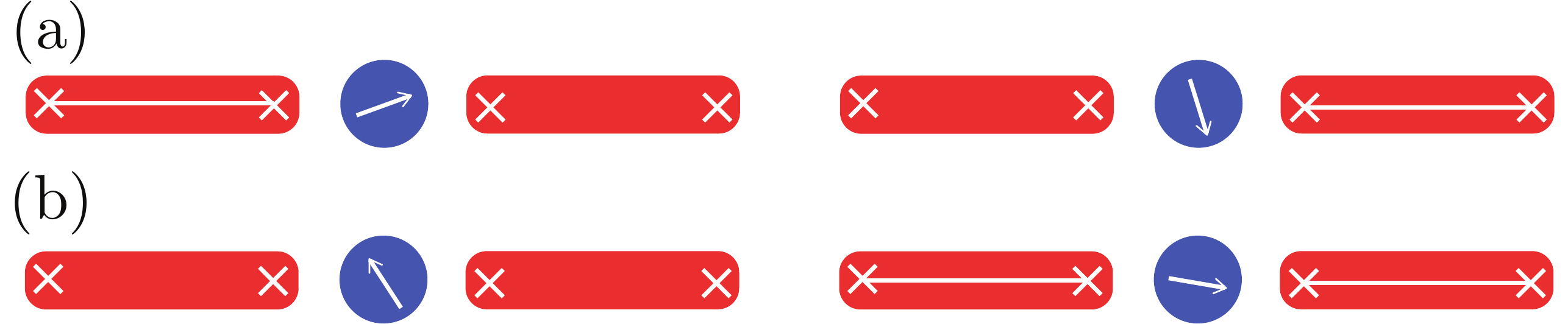}
\caption{ Schematic of the MF qubit formed by a left and right TSC (red bars) tunnel coupled by a quantum dot (blue disc) with spin-1/2 groundstate. Four MFs (white crosses) give rise to two types of MF qubit: one of odd and one of even parity.
The odd  parity of the TSCs is indicated by a straight line between the MFs. (a) The degenerate odd parity states of the MF qubit (even total system parity) with one fermion on the left TSC (left panel) and one on the right TSC (right panel). (b) The degenerate even parity states of the MF qubit (odd total system parity) with no fermions on either TSC (left panel) and with one fermion on each TSC (right panel).
}
\label{parity}
\end{figure}

Writing $\gamma_{\nu}'=f_\nu+f^\dagger_\nu$ and $\gamma_{\nu}=(f_\nu-f^\dagger_\nu)/i$ the tunneling Hamiltonian is transformed into
\begin{align}
H_T&=\sum_{\sigma\nu}d_\sigma^\dagger[it_{\nu}(f_\nu+f_\nu^\dagger)-it'_{\nu}(f_\nu-f_\nu^\dagger)]+[-it'^*_{\nu}(f_\nu-f_\nu^\dagger)-it^*_{L\nu}(f_\nu+f_\nu^\dagger)]d_\sigma\nonumber\\
&=\sum_{\sigma\nu}i(t_{\nu}'^*-t_{\nu}^*)f^\dagger_\nu d_\sigma - i(t_{\nu}'^*+t_{\nu}^*)f_{\nu} d_\sigma+i(t_{\nu}-t_{\nu}') d_\sigma^\dagger f_\nu + i(t_{\nu}'+t_{\nu}) d_\sigma^\dagger f_{\nu}^\dagger\nonumber\\
&=\sum_{\sigma\nu}it_{\nu -}^*f^\dagger_\nu d_\sigma - it_{\nu +}^*f_{\nu} d_\sigma-it_{\nu -}d_\sigma^\dagger f_\nu + it_{\nu +} d_\sigma^\dagger f_{\nu}^\dagger\,,
\end{align}
where $t_{\nu \pm}=t_{\nu}'\pm t_{\nu}$.
Using a Schrieffer-Wolff transformation, one may show that the operators $A_\nu-A_\nu^\dagger$ and $B_\nu-B_\nu^\dagger$ eliminate the tunneling Hamiltonian, $H_T=-[A_\nu-A_\nu^\dagger+B_\nu-B_\nu^\dagger,H_M+H_D]$, to first order in $t_{\nu\pm}$ where
\begin{align}
A_\nu&=i(t_{\nu}^*-t_{\nu}'^*)\sum_\sigma \left[\frac{1}{\epsilon_\sigma-2\delta_\nu} - \frac{U n_{\bar\sigma}}{(\epsilon_\sigma-2\delta_\nu)(\epsilon_\sigma+U-2\delta_\nu)} \right]f^\dagger_\nu d_\sigma\nonumber\\
&=-it_{\nu -}^*\sum_\sigma \left[\frac{1}{\epsilon_\sigma-2\delta_\nu} - \frac{U n_{\bar\sigma}}{(\epsilon_\sigma-2\delta_\nu)(\epsilon_\sigma+U-2\delta_\nu)} \right]f^\dagger_\nu d_\sigma\,,\nonumber\\
B_\nu&=i(t_{\nu}^*+t_{\nu}'^*)\sum_\sigma \left[\frac{1}{\epsilon_\sigma+2\delta_\nu} - \frac{U n_{\bar\sigma}}{(\epsilon_\sigma+2\delta_\nu)(\epsilon_\sigma+U+2\delta_\nu)} \right]f_\nu d_\sigma\nonumber\\
&=it_{\nu +}^*\sum_\sigma \left[\frac{1}{\epsilon_\sigma+2\delta_\nu} - \frac{U n_{\bar\sigma}}{(\epsilon_\sigma+2\delta_\nu)(\epsilon_\sigma+U+2\delta_\nu)} \right]f_\nu d_\sigma\,.\nonumber\\
\end{align}
We must now calculate $[A_\nu,H_T]$ and $[B_\nu,H_T]$. The commutation relations
\begin{align}
[f_\nu^\dagger d_\rho,H_T]&=i\sum_{\sigma\mu}[f_{\nu}^\dagger d_\rho,t_{-\mu}^*f^\dagger_\mu d_\sigma - t_{+\mu}^*f_{\mu} d_\sigma-t_{-\mu}d_\sigma^\dagger f_\mu + t_{+\mu} d_\sigma^\dagger f_{\mu}^\dagger]\nonumber\\
&=i\sum_{\sigma\mu}\delta_{\mu\nu}t_{+\mu}^*d_\rho d_\sigma-t_{-\mu}(\delta_{\rho\sigma}f_\nu^\dagger f_\mu-\delta_{\nu\mu}d^\dagger_\sigma d_\rho)+t_{+\mu}\delta_{\rho\sigma}f^\dagger_\nu f^\dagger_\mu\,,\nonumber\\
[f_\nu d_\rho, H_T]&=i\sum_{\sigma\mu}[f_\nu d_\rho,t_{-\mu}^*f^\dagger_\mu d_\sigma - t_{+\mu}^*f_{\mu} d_\sigma-t_{-\mu}d_\sigma^\dagger f_\mu + t_{+\mu} d_\sigma^\dagger f_{\mu}^\dagger]\nonumber\\
&=i\sum_{\sigma\mu}-t_{-\mu}^*\delta_{\mu\nu}d_\rho d_\sigma-t_{-\mu}\delta_{\rho\sigma}f_\nu f_\mu+t_{+\mu}(\delta_{\rho\sigma}f_\nu f_\mu^\dagger-\delta_{\mu\nu}d^\dagger_\sigma d_\rho)\,.
\end{align}
Note that $[ Un_{\bar\rho}f_\nu^\dagger d_\rho,H_T]= Un_{\bar\rho}[f_\nu^\dagger d_\rho,H_T]+[Un_{\bar\rho},H_T]f_\nu^\dagger d_\rho$ and
\begin{align}
[n_{\bar\rho},H_T]&=i\sum_{\sigma\nu}[n_{\bar\rho},t_{\nu -}^*f^\dagger_\nu d_\sigma - t_{\nu +}^*f_{\nu} d_\sigma-t_{\nu -}d_\sigma^\dagger f_\nu + t_{\nu +} d_\sigma^\dagger f_{\nu}^\dagger]\nonumber\\
&=i\sum_{\sigma\nu}t^*_{\nu -}\delta_{\bar\rho\sigma}d_{\bar\rho}f_\nu^\dagger-t_{\nu +}^*\delta_{\bar\rho\sigma}d_{\bar\rho}f_\nu-t_{\nu -}\delta_{\bar\rho\sigma}d_\sigma^\dagger f_\nu+t_{\nu +}\delta_{\bar\rho\sigma}d^\dagger_\sigma f_\nu^\dagger\,.
\end{align}
Taking the large on-site charging limit, $U\rightarrow \infty$, we find
\begin{align}
\sum_\nu[A_\nu,H_T]&=-i\sum_{\rho\nu} t_{\nu -}^*\left[\left(\frac{1}{\epsilon_\rho-2\delta_\nu} - \frac{ n_{\bar\rho}}{\epsilon_\rho-2\delta_\nu} \right)[f^\dagger_\nu d_\rho,H_T]-\frac{ [n_{\bar\rho},H_T]f^\dagger_\nu d_\rho}{\epsilon_\rho-2\delta_\nu}\right]\nonumber\\
&=-i\sum_{\rho\nu}\frac{t_{\nu -}^*}{\epsilon_\rho-2\delta_\nu}\left[ n_\rho[f^\dagger_\nu d_\rho,H_T]- [n_{\bar\rho},H_T]f^\dagger_\nu d_\rho\right]\nonumber\\
&=\sum_{\sigma\rho\mu\nu}\frac{t_{\nu -}^*}{\epsilon_\rho-2\delta_\nu}\left[ n_\rho(t_{+\mu}^*\delta_{\mu\nu}d_\rho d_\sigma-t_{-\mu}(\delta_{\rho\sigma}f_\nu^\dagger f_\mu-\delta_{\mu\nu}d^\dagger_\sigma d_\rho)+t_{+\mu}\delta_{\rho\sigma}f^\dagger_\nu f^\dagger_\mu)\right.\nonumber\\
&\left.- (t^*_{-\mu}\delta_{\bar\rho\sigma}d_{\bar\rho}f_\mu^\dagger-t_{+\mu}^*\delta_{\bar\rho\sigma}d_{\bar\rho}f_\mu-t_{-\mu}\delta_{\bar\rho\sigma}d_\sigma^\dagger f_\mu+t_{+\mu}\delta_{\bar\rho\sigma}d^\dagger_\sigma f_\mu^\dagger)f^\dagger_\nu d_\rho\right]\,,\nonumber\\
\sum_\nu[B_\nu,H_T]&=i\sum_{\rho\nu}t_{\nu +}^*\left[ \left(\frac{1}{\epsilon_\rho+2\delta_\nu} - \frac{ n_{\bar\rho}}{\epsilon_\rho+2\delta_\nu} \right)[f_\nu d_\rho,H_T]-\frac{ [n_{\bar\rho},H_T]f_\nu d_\rho}{\epsilon_\rho+2\delta_\nu}\right]\nonumber\\
&=i\sum_{\rho\nu}\frac{t_{\nu +}^*}{\epsilon_\rho+2\delta_\nu}\left[ n_\rho[f_\nu d_\rho,H_T]- [n_{\bar\rho},H_T]f_\nu d_\rho\right]\nonumber\\
&=-\sum_{\sigma\rho\mu\nu}\frac{t_{\nu +}^*}{\epsilon_\rho+2\delta_\nu}\left[ n_\rho(-t_{-\mu}^*\delta_{\mu\nu}d_\rho d_\sigma-t_{-\mu}\delta_{\rho\sigma}f_\nu f_\mu+t_{+\mu}(\delta_{\rho\sigma}f_\nu f_\mu^\dagger-\delta_{\mu\nu}d^\dagger_\sigma d_\rho))\right.\nonumber\\
&\left.- (t^*_{-\mu}\delta_{\bar\rho\sigma}d_{\bar\rho}f_\mu^\dagger-t_{+\mu}^*\delta_{\bar\rho\sigma}d_{\bar\rho}f_\mu-t_{-\mu}\delta_{\bar\rho\sigma}d_\sigma^\dagger f_\mu+t_{+\mu}\delta_{\bar\rho\sigma}d^\dagger_\sigma f_\mu^\dagger)f_\nu d_\rho\right]\,.\nonumber\\
\end{align}
Notice that, for $\hat O=f^\dagger_\nu,\, f_\nu$, $n_{\rho}[\hat O d_\rho,H_T]=-n_{\rho}H_T\hat O d_\rho$. The only term that survives from $H_T$ is proportional to $d_{\rho}^\dagger$ so that this term has no spin flip processes:
\begin{align}
-n_\rho H_T f_\nu^\dagger d_\rho&=i( t_{-\mu}n_\rho d^\dagger_\rho f_\mu-t_{+\mu}d^\dagger_\rho f_\mu^\dagger) f_\nu^\dagger d_\rho=i( t_{-\mu} f_\mu f_\nu^\dagger-t_{+\mu} f_\mu^\dagger f_\nu^\dagger)n_\rho\,,\nonumber\\
-n_\rho H_T f_\nu d_\rho&=-i(-t_{+\mu}n_\rho d^\dagger_\rho f_\mu^\dagger+t_{-\mu}d_\rho^\dagger f_\mu)  f_\nu d_\rho=-i(-t_{+\mu}f_\mu^\dagger f_\nu +t_{-\mu} f_\mu f_\nu)n_\rho\,.
\end{align}
Therefore, these terms do not involve spin flips and
\begin{align}
\sum_\nu[A_\nu,H_T]&=-i\sum_{\rho\nu}t_{\nu -}^*\left[ \left(\frac{1}{\epsilon_\rho-2\delta_\nu} - \frac{ n_{\bar\rho}}{\epsilon_\rho-2\delta_\nu} \right)[f^\dagger_\nu d_\rho,H_T]-\frac{ [n_{\bar\rho},H_T]f^\dagger_\nu d_\rho}{\epsilon_\rho-2\delta_\nu}\right]\nonumber\\
&=-i\sum_{\rho\nu}\frac{t_{\nu -}^*}{\epsilon_\rho-2\delta_\nu}\left[ n_\rho[f^\dagger_\nu d_\rho,H_T]- [n_{\bar\rho},H_T]f^\dagger_\nu d_\rho\right]\nonumber\\
&=\sum_{\sigma\rho\mu\nu}\frac{t_{\nu -}^*}{\epsilon_\rho-2\delta_\nu}\left[ (t_{-\mu}f_\mu f_\nu^\dagger-t_{+\mu}f_\mu^\dagger f_\nu^\dagger)\delta_{\sigma\rho}n_\rho - (t^*_{-\mu}\delta_{\bar\rho\sigma}d_{\bar\rho}f_\mu^\dagger-t_{+\mu}^*\delta_{\bar\rho\sigma}d_{\bar\rho}f_\mu-t_{-\mu}\delta_{\bar\rho\sigma}d_\sigma^\dagger f_\mu+t_{+\mu}\delta_{\bar\rho\sigma}d^\dagger_\sigma f_\mu^\dagger)f^\dagger_\nu d_\rho\right]\nonumber\\
&=\sum_{\sigma\rho\mu\nu}\frac{t_{\nu -}^*}{\epsilon_\rho-2\delta_\nu}\left[ t_{-\mu}\delta_{\sigma\rho}n_\rho f_\mu f_\nu^\dagger-t_{+\mu}\delta_{\sigma\rho}n_\rho f_\mu^\dagger f_\nu^\dagger- (-t_{-\mu}\delta_{\bar\rho\sigma}d_\sigma^\dagger f_\mu+t_{+\mu}\delta_{\bar\rho\sigma}d^\dagger_\sigma f_\mu^\dagger)f^\dagger_\nu d_\rho\right]\,,\nonumber\\
\sum_\nu[B_\nu,H_T]&=i\sum_{\rho\nu} t_{\nu +}^*\left[\left(\frac{1}{\epsilon_\rho+2\delta_\nu} - \frac{ n_{\bar\rho}}{\epsilon_\rho+2\delta_\nu} \right)[f_\nu d_\rho,H_T]-\frac{ [n_{\bar\rho},H_T]f_\nu d_\rho}{\epsilon_\rho+2\delta_\nu}\right]\nonumber\\
&=i\sum_{\rho\nu}\frac{t_{\nu +}^*}{\epsilon_\rho+2\delta_\nu}\left[ n_\rho[f_\nu d_\rho,H_T]- [n_{\bar\rho},H_T]f_\nu d_\rho\right]\nonumber\\
&=-\sum_{\sigma\rho\mu\nu}\frac{t_{\nu +}^*}{\epsilon_\rho+2\delta_\nu}[-t_{+\mu}\delta_{\sigma\rho}n_\rho f_\mu^\dagger f_\nu+t_{-\mu}\delta_{\sigma\rho}n_\rho f_\mu f_\nu- (t^*_{-\mu}\delta_{\bar\rho\sigma}d_{\bar\rho}f_\mu^\dagger-t_{+\mu}^*\delta_{\bar\rho\sigma}d_{\bar\rho}f_\mu-t_{-\mu}\delta_{\bar\rho\sigma}d_\sigma^\dagger f_\mu+t_{+\mu}\delta_{\bar\rho\sigma}d^\dagger_\sigma f_\mu^\dagger)\nonumber\\
&\,\,\,\,\,\,\times f_\nu d_\rho]\nonumber\\
&=-\sum_{\sigma\rho\mu\nu}\frac{t_{\nu +}^*}{\epsilon_\rho+2\delta_\nu}\left[-t_{+\mu}\delta_{\sigma\rho}n_\rho f_\mu^\dagger f_\nu+t_{-\mu}\delta_{\sigma\rho}n_\rho f_\mu f_\nu- (-t_{-\mu}\delta_{\bar\rho\sigma}d_\sigma^\dagger f_\mu+t_{+\mu}\delta_{\bar\rho\sigma}d^\dagger_\sigma f_\mu^\dagger)f_\nu d_\rho\right]\,.
\end{align}
Let us consider processes when only one TSC is involved in then tunneling, $\mu=\nu$ [Fig.~\ref{virt}(a),~(b)]:
\begin{align}
\sum_\nu[A_\nu,H_T]&=\sum_{\sigma\rho\mu\nu}\frac{t_{\nu -}^*}{\epsilon_\rho-2\delta_\nu}\left[ t_{-\mu}\delta_{\sigma\rho}n_\rho f_\mu f_\nu^\dagger- (t^*_{-\mu}\delta_{\bar\rho\sigma}d_{\bar\rho}f_\mu^\dagger-t_{+\mu}^*\delta_{\bar\rho\sigma}d_{\bar\rho}f_\mu-t_{-\mu}\delta_{\bar\rho\sigma}d_\sigma^\dagger f_\mu+t_{+\mu}\delta_{\bar\rho\sigma}d^\dagger_\sigma f_\mu^\dagger)f^\dagger_\nu d_\rho\right]\nonumber\\
&=\sum_{\rho\nu}\frac{t_{-}^*}{\epsilon_\rho-2\delta_\nu}\left[ t_{-}n_\rho f_\nu f_\nu^\dagger- (-t_{+}^*d_{\bar\rho}f_\nu-t_{-}d_{\bar\rho}^\dagger f_\nu)f_\nu^\dagger d_\rho\right]\nonumber\\
&=\sum_{\rho\nu}\frac{t_{-}^*}{\epsilon_\rho-2\delta}\left[ t_{-}n_\rho f_\nu f_\nu^\dagger+t_{-}d_{\bar\rho}^\dagger f_\nu f_\nu^\dagger d_\rho\right]\nonumber\\
\sum_\nu[B_\nu,H_T]&=-\sum_{\sigma\rho\mu\nu}\frac{t_{\nu +}^*}{\epsilon_\rho+2\delta_\nu}\left[-t_{+\mu}\delta_{\sigma\rho}n_\rho f_\mu^\dagger f_\nu- (t^*_{-\mu}\delta_{\bar\rho\sigma}d_{\bar\rho}f_\mu^\dagger-t_{+\mu}^*\delta_{\bar\rho\sigma}d_{\bar\rho}f_\mu-t_{-\mu}\delta_{\bar\rho\sigma}d_\sigma^\dagger f_\mu+t_{+\mu}\delta_{\bar\rho\sigma}d^\dagger_\sigma f_\mu^\dagger)f_\nu d_\rho\right]\nonumber\\
&=-\sum_{\rho\nu}\frac{t_{+}^*}{\epsilon_\rho+2\delta}\left[-t_{+}n_\rho f_\nu^\dagger f_\nu - (t^*_{-}d_{\bar\rho}f_\nu^\dagger+t_{+}d^\dagger_{\bar\rho} f_\nu^\dagger)f_\nu d_\rho\right]\nonumber\\
&=\sum_{\rho\nu}\frac{t_{+}^*}{\epsilon_\rho+2\delta}\left[t_{+}n_\rho f_\nu^\dagger f_\nu +t_{+}d^\dagger_{\bar\rho} f_\nu^\dagger f_\nu d_\rho\right]\,,
\end{align}
where the final inequalities for each term is due to the single occupancy of the dot. Summing these together, with their Hermitian conjugate, we get
\eq{
\mathcal H_s=\sum_{\rho\nu}\left(\frac{|t_{\nu+}|^2}{\epsilon_\rho+2\delta_\nu}f_\nu^\dagger f_\nu +\frac{|t_{\nu-}|^2}{\epsilon_\rho-2\delta_\nu}f_\nu f_\nu^\dagger\right)\left(2n_\rho+d^\dagger_{\bar\rho}d_\rho+d^\dagger_{\rho}d_{\bar\rho}\right)\,.
}
Processes involving multiple TSCs, $\mu=\bar\nu$, are calculated from
\begin{align}
\sum_\nu[A_\nu,H_T]&=\sum_{\sigma\rho\mu\nu}\frac{t_{\nu -}^*}{\epsilon_\rho-2\delta_\nu}\left[ t_{-\mu}\delta_{\sigma\rho}n_\rho f_\mu f_\nu^\dagger-t_{+\mu}\delta_{\sigma\rho}n_\rho f_\mu^\dagger f_\nu^\dagger- (t^*_{-\mu}\delta_{\bar\rho\sigma}d_{\bar\rho}f_\mu^\dagger-t_{+\mu}^*\delta_{\bar\rho\sigma}d_{\bar\rho}f_\mu-t_{-\mu}\delta_{\bar\rho\sigma}d_\sigma^\dagger f_\mu+t_{+\mu}\delta_{\bar\rho\sigma}d^\dagger_\sigma f_\mu^\dagger)f^\dagger_\nu d_\rho\right]\nonumber\\
&=\sum_{\sigma\rho\mu\nu}\frac{t_{\nu -}^*}{\epsilon_\rho-2\delta_\nu}\left[ t_{-\mu}\delta_{\sigma\rho}n_\rho f_\mu f_\nu^\dagger-t_{+\mu}\delta_{\sigma\rho}n_\rho f_\mu^\dagger f_\nu^\dagger- (-t_{-\mu}\delta_{\bar\rho\sigma}d_\sigma^\dagger f_\mu+t_{+\mu}\delta_{\bar\rho\sigma}d^\dagger_\sigma f_\mu^\dagger)f^\dagger_\nu d_\rho\right]\,,\nonumber\\
\sum_\nu[B_\nu,H_T]&=-\sum_{\sigma\rho\mu\nu}\frac{t_{\nu +}^*}{\epsilon_\rho+2\delta_\nu}[-t_{+\mu}\delta_{\sigma\rho}n_\rho f_\mu^\dagger f_\nu+t_{-\mu}\delta_{\sigma\rho}n_\rho f_\mu f_\nu- (t^*_{-\mu}\delta_{\bar\rho\sigma}d_{\bar\rho}f_\mu^\dagger-t_{+\mu}^*\delta_{\bar\rho\sigma}d_{\bar\rho}f_\mu-t_{-\mu}\delta_{\bar\rho\sigma}d_\sigma^\dagger f_\mu+t_{+\mu}\delta_{\bar\rho\sigma}d^\dagger_\sigma f_\mu^\dagger)\nonumber\\
&\,\,\,\,\,\,\times f_\nu d_\rho]\nonumber\\
&=-\sum_{\sigma\rho\mu\nu}\frac{t_{\nu +}^*}{\epsilon_\rho+2\delta_\nu}\left[-t_{+\mu}\delta_{\sigma\rho}n_\rho f_\mu^\dagger f_\nu+t_{-\mu}\delta_{\sigma\rho}n_\rho f_\mu f_\nu- (-t_{-\mu}\delta_{\bar\rho\sigma}d_\sigma^\dagger f_\mu+t_{+\mu}\delta_{\bar\rho\sigma}d^\dagger_\sigma f_\mu^\dagger)f_\nu d_\rho\right]\,.
\end{align}
Because we will have to add the Hermitian conjugates of these terms, notice that 
\begin{align}
\left[\sum_{\rho\nu}\frac{t_{\bar\nu -}^*}{\epsilon_\rho-2\delta_{\bar\nu}} t_{\nu -}n_\rho f_\nu f_{\bar\nu}^\dagger\right]^\dagger&=\sum_{\rho\nu}\frac{t_{\bar\nu -}^*}{\epsilon_\rho-2\delta_{\nu}} t_{\nu -}n_\rho f_\nu f_{\bar\nu}^\dagger\,,\nonumber\\
\left[\sum_{\rho\nu}\frac{t_{\bar\nu +}^*}{\epsilon_\rho+2\delta_{\bar\nu}} t_{\nu +}n_\rho f_\nu^\dagger f_{\bar\nu}\right]^\dagger&=\sum_{\rho\nu}\frac{t_{\bar\nu +}^*}{\epsilon_\rho+2\delta_{\nu}} t_{\nu +}n_\rho f_\nu^\dagger f_{\bar\nu}\,,\nonumber\\
\left[\sum_{\rho\nu}\frac{t_{\bar\nu -}^*}{\epsilon_\rho-2\delta_{\bar\nu}} t_{\nu -}d_{\bar\rho}^\dagger d_\rho f_\nu f_{\bar\nu}^\dagger\right]^\dagger&=\sum_{\rho\nu}\frac{t_{\bar\nu -}^*}{\epsilon_{\bar\rho}-2\delta_{\nu}} t_{\nu -}d_{\bar\rho}^\dagger d_\rho  f_\nu f_{\bar\nu}^\dagger\,,\nonumber\\
\left[\sum_{\rho\nu}\frac{t_{\bar\nu +}^*}{\epsilon_\rho+2\delta_{\bar\nu}} t_{\nu +}d_{\bar\rho}^\dagger d_\rho f_\nu^\dagger f_{\bar\nu}\right]^\dagger&=\sum_{\rho\nu}\frac{t_{\bar\nu +}^*}{\epsilon_{\bar\rho}+2\delta_{\nu}} t_{\nu +}d_{\bar\rho}^\dagger d_\rho f_\nu^\dagger f_{\bar\nu}\,,\nonumber\\
\end{align}
so that the contribution from the transfer of the fermions [Fig.~\ref{virt}(c),~(d)] is
\begin{align}
\mathcal H_o=\sum_{\rho\nu}&\left[\left(\frac{1}{{\epsilon_\rho-2\delta_{\bar\nu}}}+\frac{1}{{\epsilon_\rho-2\delta_{\nu}}}\right)t_{\nu -}t_{\bar\nu -}^*f_\nu f^\dagger_{\bar\nu}+\left(\frac{1}{{\epsilon_\rho+2\delta_{\bar\nu}}}+\frac{1}{{\epsilon_\rho+2\delta_{\nu}}}\right)t_{\nu +}t_{\bar\nu +}^*f^\dagger_\nu f_{\bar\nu}\right]n_\rho\nonumber\\
&+\left[\left(\frac{1}{\epsilon_\rho-2\delta_{\bar\nu}}+\frac{1}{\epsilon_{\bar\rho}-2\delta_{\nu}}\right)t^*_{\bar\nu -}t_{\nu -}f_\nu f_{\bar\nu}^\dagger+\left(\frac{1}{\epsilon_\rho+2\delta_{\bar\nu}}+\frac{1}{\epsilon_{\bar\rho}+2\delta_{\nu}}\right)t^*_{\bar\nu +}t_{\nu +} f_\nu^\dagger f_{\bar\nu}\right]d^\dagger_{\bar\rho}d_\rho\nonumber\\
=\sum_{\rho\nu}&\left(\frac{n_\rho+d^\dagger_{\bar\rho}d_\rho}{\epsilon_\rho-2\delta_{\bar\nu}}+\frac{n_\rho+d_\rho^\dagger d_{\bar\rho}}{{\epsilon_\rho-2\delta_{\nu}}}\right)t^*_{\bar\nu -}t_{\nu -}f_\nu f_{\bar\nu}^\dagger+\left(\frac{n_\rho+d^\dagger_{\bar\rho}d_\rho}{\epsilon_\rho+2\delta_{\bar\nu}}+\frac{n_\rho+d_\rho^\dagger d_{\bar\rho}}{\epsilon_{\rho}+2\delta_{\nu}}\right)t^*_{\bar\nu +}t_{\nu +} f_\nu^\dagger f_{\bar\nu}\,.
\end{align}

Next notice that 
\begin{align}
\left(-\sum_{\rho\nu}\frac{t_{\bar\nu -}^*t_{\nu +}}{\epsilon_\rho-2\delta_{\bar\nu}}d^\dagger_{\bar\rho} f_\nu^\dagger f^\dagger_{\bar\nu} d_\rho\right)^\dagger&=-\sum_{\rho\nu}\frac{t_{\bar\nu +}^*t_{\nu -}}{\epsilon_{\bar\rho}-2\delta_{\nu}}d_{\bar\rho}^\dagger f_\nu f_{\bar\nu} d_\rho\nonumber\,,\\
\left(-\sum_{\rho\nu}\frac{t_{\bar\nu +}^*t_{\nu -}}{\epsilon_\rho+2\delta_{\bar\nu}}d_{\bar\rho}^\dagger f_\nu f_{\bar\nu} d_\rho\right)^\dagger&=-\sum_{\rho\nu}\frac{t_{\bar\nu -}^*t_{\nu +}}{\epsilon_{\bar\rho}+2\delta_{\nu}}d_{\bar\rho}^\dagger f_\nu^\dagger f_{\bar\nu}^\dagger d_\rho\,,
\end{align}
so that terms acting the even parity sector [Fig.~\ref{virt}(e),~(f)] are
\eq{
\mathcal H_e=-\sum_{\rho\nu}t_{\bar\nu +}^*t_{\nu -}\left(\frac{n_\rho+d_{\bar\rho}^\dagger d_\rho}{\epsilon_\rho+2\delta_{\bar\nu}}+\frac{n_{\rho}+d_{\rho}^\dagger d_{\bar\rho}}{\epsilon_{\rho}-2\delta_{\nu}}\right)  f_\nu f_{\bar\nu} + t_{\bar\nu -}^*t_{\nu +}\left(\frac{n_\rho+d_{\bar\rho}^\dagger d_\rho}{\epsilon_\rho-2\delta_{\bar\nu}}+\frac{n_\rho+d_{\rho}^\dagger d_{\bar\rho}}{\epsilon_{\rho}+2\delta_{\nu}}\right)  f_\nu^\dagger f_{\bar\nu}^\dagger\,.
}

Summing up the results we have the tunneling Hamiltonian to second order in $t_{\nu\pm}$:
\begin{align}
\mathcal  H_T&=\mathcal H_s+\mathcal H_e+\mathcal H_o\,,\nonumber\\
\mathcal H_s&=\sum_{\sigma\nu}\left(\frac{|t_{\nu +}|^2}{\epsilon_\sigma+2\delta_{\nu}}f_\nu^\dagger f_\nu +\frac{|t_{\nu -}|^2}{\epsilon_\sigma-2\delta_{\nu}}f_\nu f_\nu^\dagger\right)\left(2n_\sigma+d^\dagger_{\bar\sigma}d_\sigma+d^\dagger_{\sigma}d_{\bar\sigma}\right)\,,\nonumber\\
\mathcal H_o&=\sum_{\sigma\nu}\left(\frac{n_\sigma+d^\dagger_{\bar\sigma}d_\sigma}{\epsilon_\sigma-2\delta_{\bar\nu}}+\frac{n_\sigma+d_\sigma^\dagger d_{\bar\sigma}}{{\epsilon_\sigma-2\delta_{\nu}}}\right)t^*_{\bar\nu -}t_{\nu -}f_\nu f_{\bar\nu}^\dagger+\left(\frac{n_\sigma+d^\dagger_{\bar\sigma}d_\sigma}{\epsilon_\sigma+2\delta_{\bar\nu}}+\frac{n_\sigma+d_\sigma^\dagger d_{\bar\sigma}}{\epsilon_{\sigma}+2\delta_{\nu}}\right)t^*_{\bar\nu +}t_{\nu +} f_\nu^\dagger f_{\bar\nu}\,,\nonumber\\
\mathcal H_e&=-\sum_{\sigma\nu}t_{\bar\nu +}^*t_{\nu -}\left(\frac{n_\sigma+d_{\bar\sigma}^\dagger d_\sigma}{\epsilon_\sigma+2\delta_{\bar\nu}}+\frac{n_\sigma+d_{\sigma}^\dagger d_{\bar\sigma}}{\epsilon_{\sigma}-2\delta_{\nu}}\right)  f_\nu f_{\bar\nu} + t_{\bar\nu -}^*t_{\nu +}\left(\frac{n_\sigma+d_{\bar\sigma}^\dagger d_\sigma}{\epsilon_\sigma-2\delta_{\bar\nu}}+\frac{n_\sigma+d_{\sigma}^\dagger d_{\bar\sigma}}{\epsilon_{\sigma}+2\delta_{\nu}}\right)  f_\nu^\dagger f_{\bar\nu}^\dagger\,.
\end{align}
When the splitting is equal in both TSCs $\delta_\nu=\delta$, we obtain Eq.~(\ref{Hsw}) in the main text,
\begin{align}
\mathcal H_T&=\mathcal H_s+\mathcal H_o+\mathcal H_e\,,\nonumber\\
\mathcal H_s&=\sum_{\sigma,\nu}\left(\frac{| t_{\nu -}|^2}{\epsilon_\sigma-2\delta}f_\nu f_\nu^\dagger+\frac{| t_{\nu +}|^2}{\epsilon_\sigma+2\delta}f_\nu^\dagger f_\nu\right)\left(2n_\sigma+d^\dagger_{\bar\sigma}d_\sigma+d^\dagger_{\sigma}d_{\bar\sigma}\right)\,,\nonumber\\
\mathcal H_o&=\sum_{\sigma,\nu}\left(\frac{ t^*_{\bar\nu -} t_{\nu -}}{\epsilon_\sigma-2\delta} f_\nu f_{\bar\nu}^\dagger+\frac{ t^*_{\bar\nu +} t_{\nu +}}{\epsilon_{\sigma}+2\delta} f_\nu^\dagger f_{\bar\nu}\right)(2n_\sigma+d^\dagger_{\bar\sigma}d_\sigma+d^\dagger_{\sigma}d_{\bar\sigma})\,,\nonumber\\
\mathcal H_e&=-\sum_{\sigma,\nu} t_{\bar\nu -}^* t_{\nu +}\left(\frac{n_\sigma+d_{\bar\sigma}^\dagger d_\sigma}{\epsilon_\sigma-2\delta}+\frac{n_\sigma+d_{\sigma}^\dagger d_{\bar\sigma}}{\epsilon_{\sigma}+2\delta}\right)  f_\nu^\dagger f_{\bar\nu}^\dagger+t_{\bar\nu +}^* t_{\nu -}\left(\frac{n_\sigma+d_{\bar\sigma}^\dagger d_\sigma}{\epsilon_\sigma+2\delta}+\frac{n_\sigma+d_{\sigma}^\dagger d_{\bar\sigma}}{\epsilon_{\sigma}-2\delta}\right)  f_\nu f_{\bar\nu}\,.
\end{align}

\section{Full Exchange Hamiltonian}

The full interaction between the MF qubit and the spin qubit can be written down as the exchange Hamiltonian
\eq{
\mathcal  H_T=\sum_{\kappa,\lambda=0,\ldots,4}J_{\kappa\lambda}\sigma_\kappa\eta_\lambda\,,
}
in which
\eq{
J_{\kappa\lambda}=\left(
\begin{array}{cccc}
B_1 & B_2 & B_3 & B_4\\
B_1 & B_2 & B_3 & B_4\\
0 & 0 & 0 & 0\\
B_5 & B_6 & B_7 & B_8\\
\end{array}
\right)\,,
\label{J}
}
where
\begin{align}
B_1&=(\mathcal{C}_{++}+\mathcal{C}_{-+})(\Gamma_{+r}+\Gamma_{+l})+(\mathcal{C}_{+-}+\mathcal{C}_{--})(\Gamma_{-r}+\Gamma_{-l})\,,\nonumber\\
B_2&=(\mathcal{C}_{++}+\mathcal{C}_{-+})(\tilde\Gamma_{+r}+\tilde\Gamma_{+l})-(\mathcal{C}_{+-}+\mathcal{C}_{--})(\tilde\Gamma_{-r}+\tilde\Gamma_{-l})\,,\nonumber\\
B_3&=i\left[(\mathcal{C}_{++}+\mathcal{C}_{-+})(\tilde\Gamma_{+r}-\tilde\Gamma_{+l})+(\mathcal{C}_{+-}+\mathcal{C}_{--})(\tilde\Gamma_{-r}-\tilde\Gamma_{-l})\right]\,,\nonumber\\
B_4&=(\mathcal{C}_{++}+\mathcal{C}_{-+})(\Gamma_{+r}-\Gamma_{+l})-(\mathcal{C}_{+-}+\mathcal{C}_{--})(\Gamma_{-r}-\Gamma_{-l})\,,\nonumber\\
B_5&=(\mathcal{C}_{++}-\mathcal{C}_{-+})(\Gamma_{+r}+\Gamma_{+l})+(\mathcal{C}_{+-}-\mathcal{C}_{--})(\Gamma_{-r}+\Gamma_{-l})\,,\nonumber\\
B_6&=(\mathcal{C}_{++}-\mathcal{C}_{-+})(\tilde\Gamma_{+r}+\tilde\Gamma_{+l})-(\mathcal{C}_{+-}-\mathcal{C}_{--})(\tilde\Gamma_{-r}+\tilde\Gamma_{-l})\,,\nonumber\\
B_7&=i\left[(\mathcal{C}_{++}-\mathcal{C}_{-+})(\tilde\Gamma_{+r}-\tilde\Gamma_{+l})+(\mathcal{C}_{+-}-\mathcal{C}_{--})(\tilde\Gamma_{-r}-\tilde\Gamma_{-l})\right]\,,\nonumber\\
B_8&=(\mathcal{C}_{++}-\mathcal{C}_{-+})(\Gamma_{+r}-\Gamma_{+l})-(\mathcal{C}_{+-}-\mathcal{C}_{--})(\Gamma_{-r}-\Gamma_{-l})\,.
\end{align}
Here, $\Gamma_{\pm\nu}=|t_{\nu \pm}|^2$ and $\tilde\Gamma_{\pm\nu}=t^*_{\pm\bar\nu}t_{\nu \pm}$ and $\mathcal{C}_{\sigma\rho}=1/(\epsilon_\sigma+2\rho\delta)$ or, with a $\sigma$ and $\rho$ independent denominator,
\eq{
\mathcal{C}_{\sigma\rho}=\frac{(\epsilon_0-2\rho\delta)(\epsilon_0^2+\Delta^2-4\delta^2)-2\epsilon_0\Delta^2+\sigma\Delta[2\epsilon_0(\epsilon_0-2\rho\delta)-(\epsilon_0^2+\Delta^2-4\delta^2)]}{[(\epsilon_0-\Delta)^2-4\delta^2][(\epsilon_0+\Delta)^2-4\delta^2]}\,,
}
where we have written $\epsilon_\sigma=\epsilon_0+\sigma\Delta$.

We consider the limit that the length of the TSCs is infinite and the dot is placed between them, so that $t_{l}'=t_{r}=0$ and $\delta=0$, thus $t_{\pm r}=\pm t_{r}'$ and $t_{\pm l}= t_{l}$. When the difference in phase between $t_r'$ and $t_l$ is $\phi$, we find that $\Gamma_{\pm r}=|t_r'|^2$, $\Gamma_{\pm l}=|t_l|^2$, $\Gamma_{\pm r}=\pm t_r' t_l e^{i\phi}$, and $\Gamma_{\pm l}=\pm t_r' t_l e^{-i\phi}$. The exchange interaction becomes
\begin{align}
B_1&=\mathcal{D}^+(|t_r'|^2+t^2_l)\,,\nonumber\\
B_2&=2\mathcal{D}^+ t_r' t_l \cos\phi\nonumber\\
B_3&=0\,,\nonumber\\
B_4&=0\,,\nonumber\\
B_5&=\mathcal{D}^-(|t_r'|^2+t_l^2)\,,\nonumber\\
B_6&=2\mathcal{D}^- t_r' t_l \cos\phi\,,\nonumber\\
B_7&=0\,,\nonumber\\
B_8&=0\,,
\end{align}
where
\begin{align}
\mathcal{D}^+ &= \mathcal{C}_{++}+\mathcal{C}_{+-}+\mathcal{C}_{-+}+\mathcal{C}_{-+}=\frac{\epsilon_0}{\epsilon_0^2-\Delta^2}\,,\nonumber\\
\mathcal{D}^- &= \mathcal{C}_{++}+\mathcal{C}_{+-}-\mathcal{C}_{-+}-\mathcal{C}_{-+}=\frac{\Delta}{\epsilon_0^2-\Delta^2}\,.
\end{align}
When $\Delta=0$ and $t_r'=t_l=t$, this reduces to Eq.~(\ref{HCP}).

\section{Hybrid CNOT Gate}
Let us introduce the  hCP gate $U'_{\textrm{hCP}}=\exp[i\pi(1-\sigma_3)(1-\eta_3)/4]=(1+\sigma_3+\eta_3 -\sigma_3 \eta_3)/2$
 and relate it to the one used in the main text,
 $U_{\textrm{hCP}}=\exp[i\pi(1+\sigma_3)(1+\eta_3)/4]=(1-\sigma_3-\eta_3 -\sigma_3 \eta_3)/2$. Note that $U'_{\textrm{hCP}}$ reduces to the `canonical form' of the conditional phase gate for identical qubit types.
 Next, we note that $U'_{\textrm{hCP}}=U_{\textrm{hCP}} R_{SQ}(-\pi) R_{MQ}(-\pi)$, where $R_{SQ}(\phi)=\exp[i\phi \sigma_3 /2]$ and $R_{MQ}(\phi)=\exp[i\phi \eta_3 /2] $ are the phase gates on the spin and MF qubit, respectively.
Then, we get the corresponding hybrid CNOT gate $U_{\textrm{hCNOT}}^{'31} =(1+\sigma_3+\eta_1 -\sigma_3 \eta_1)/2$
from $U'_{\textrm{hCP}}$ by a Hadamard operation $H_{MQ}=(\eta_1 +\eta_3)/\sqrt{2}$ (which takes $\eta_3$ into $\eta_1$), $U_{\textrm{hCNOT}}^{'31}=H_{MQ} U'_{\textrm{hCP}} H_{MQ}$, and thus \begin{equation}
U_{\textrm{hCNOT}}^{'31}= H_{MQ} U'_{\textrm{hCP}} R_{SQ}(-\pi) R_{MQ}(-\pi) H_{MQ}=U_{\textrm{hCNOT}}^{31} R_{SQ}(-\pi)  H_{MQ} R_{MQ}(-\pi) H_{MQ},
\end{equation}
where $U_{\textrm{hCNOT}}^{31}=(1-\sigma_3-\eta_1 -\sigma_3 \eta_1)/2$ (used in the main text).
Thus, we can get the `canonical form' of the CNOT gate, $U_{\textrm{hCNOT}}^{'31}$, from $U_{\textrm{hCNOT}}^{31}$ by simple single-qubit unitary operations. And similarly for $U_{\textrm{hCNOT}}^{13}$.
Note that the phase gate $R_{MQ}(-\pi)$ can be obtained by braiding since it is the square of the $\pi/4$ phase gate.

\section{Effective Interaction between Majorana Fermion Qubits}
In this section we derive an effective Hamiltonian for the interaction of neighboring
hybrid qubits, labeled $(1)$ and $(2)$, in a MaSH network [Fig.~\ref{setup}(b)]. We assume that adjacent spin qubits couple via an isotropic exchange 
interaction of the form 
\begin{equation}
\mathcal{H}^{(12)}_{SQ}
=
\mathcal J
\left[
\sigma^{(1)}_{1}
\sigma^{(2)}_{1}
+
\sigma^{(1)}_{2}
\sigma^{(2)}_{2}
+
\sigma^{(1)}_{3}
\sigma^{(2)}_{3}
\right]
\end{equation}
and that the MF qubits couple to the spin qubits via 
\begin{equation}
\mathcal{H}_{\textrm{hCP}}
=
\frac{2|t|^{2}}{\epsilon_0} 
\left[ 
\mathbb 1+
\sigma^{(1)}_{1}
+
\eta^{(1)}_{1}
+
\sigma^{(1)}_{1}
\eta^{(1)}_{1}
+
\sigma^{(2)}_{1}
+
\eta^{(2)}_{1}
+
\sigma^{(2)}_{1}
\eta^{(2)}_{1}
\right]\,,
\end{equation}
according to Eq.~(\ref{HCP}) in the main text. When $\mathcal J\gg2|t|^{2}/\epsilon_{0}$,
we can make a Schrieffer-Wolff transformation on $\mathcal H^{(12)}_{SQ}$, using $\mathcal H_{\textrm{hCP}}$ as a perturbation, which gives an effective coupling between two hybrid qubits
 up to second order in $|t|^{2}/\epsilon_{0}$,
\begin{equation}
\mathcal{H}^{(12)}_{HQ}
=
\mathcal{H}^{(12)}_{SQ}+\mathcal{H}^{(12)}_{MQ}\,,
\end{equation}
where
\begin{equation}
\mathcal{H}^{(12)}_{MQ}=-
\lim_{\varepsilon\rightarrow0^{+}}
\frac{i}{2\hbar}
\int_{0}^{\infty} \mathrm{d}\tau \  e^{-\varepsilon\tau}
\left[
\mathcal{H}_{\textrm{hCP}}(\tau),\mathcal{H}_{\textrm{hCP}}
\right]\,.
\label{MQMQ}
\end{equation}
Here, $\mathcal{H}_{\textrm{hCP}}(\tau)$ is the time-evolution of $\mathcal{H}_{\textrm{hCP}}$ under the unperturbed Hamiltonian $\mathcal{H}^{(12)}_{SQ}$, 
\begin{equation}
\begin{split}
\mathcal{H}_{\textrm{hCP}}(\tau)
&=e^{i\mathcal{H}^{(12)}_{SQ}\tau/\hbar}
\mathcal{H}_{\textrm{hCP}}
e^{-i\mathcal{H}^{(12)}_{SQ}\tau/\hbar}
\\
&=
\left[
e^{i\omega_\mathcal{J}\sigma^{(1)}_{1}
\sigma^{(2)}_{1}\tau}
e^{i\omega_\mathcal{J}
\sigma^{(1)}_{2}
\sigma^{(2)}_{2}
\tau}
e^{i\omega_\mathcal{J}
\sigma^{(1)}_{3}
\sigma^{(2)}_{3}\tau}
\right]
\mathcal{H}_{\textrm{hCP}}
\left[
e^{-i\omega_\mathcal{J}\sigma^{(1)}_{1}
\sigma^{(2)}_{1}\tau}
e^{-i\omega_\mathcal{J}
\sigma^{(1)}_{2}
\sigma^{(2)}_{2}
\tau}
e^{-i\omega_\mathcal{J}
\sigma^{(1)}_{3}
\sigma^{(2)}_{3}\tau}
\right]
\\
&=
\frac{2|t|^{2}}{\epsilon_0} 
\left\{\mathbb 1+
\eta^{(1)}_{1}+\eta^{(2)}_{1}
+
e^{2i\omega_\mathcal{J}
\sigma^{(1)}_{2}
\sigma^{(2)}_{2}
\tau}
e^{2i\omega_\mathcal{J}
\sigma^{(1)}_{3}
\sigma^{(2)}_{3}\tau}
\left[
\sigma^{(1)}_{1}
+
\sigma^{(1)}_{1}
\eta^{(1)}_{1}
+
\sigma^{(2)}_{1}
+
\sigma^{(2)}_{1}
\eta^{(2)}_{1}
\right]
\right\}\,,
\end{split}
\end{equation}
with $\omega_{\mathcal J}=\mathcal J/\hbar$. Evaluating the commutator in Eq.~(\ref{MQMQ})
\begin{equation}
\begin{split}
\left[
\mathcal{H}_{\textrm{hCP}}(\tau),\mathcal{H}_{\textrm{hCP}}
\right]
=
4i
\left(\frac{2|t|^{2}}{\epsilon_0} \right)^{2}
\cos(2\omega_\mathcal{J} \tau)\sin(2\omega_\mathcal{J} \tau)
\left[\sigma^{(1)}_{2}\sigma^{(2)}_{2}+\sigma^{(1)}_{3}\sigma^{(2)}_{3}\right]
\left[1-\eta^{(1)}_{1}\eta^{(2)}_{1}\right]
\end{split}
\end{equation}
and using the integral
\begin{equation}
\begin{split}
\lim_{\varepsilon\rightarrow0^{+}}
\int_{0}^{\infty}\mathrm{d}\tau \ e^{-\varepsilon \tau}    \sin(2\omega_\mathcal{J} \tau)\cos(2\omega_\mathcal{J} \tau)
&=
\frac{1}{8\omega_\mathcal{J}}\,,
\end{split}
\end{equation}
we find
\begin{equation}
\begin{split}
\mathcal H^{(12)}_{MQ}=
\frac{|t|^{4}}{\epsilon^{2}_0\mathcal J}
\left[\sigma^{(1)}_{2}\sigma^{(2)}_{2}+\sigma^{(1)}_{3}\sigma^{(2)}_{3}\right]
\left[1-\eta^{(1)}_{1}\eta^{(2)}_{1}\right]\,,
\end{split}
\end{equation}
an effective exchange coupling between adjacent MF qubits which is modulated by the corresponding spin qubits. Applying this interaction for a time $\tau_{MF}=\pi\hbar\epsilon_0\mathcal J/|t|^4$, we obtain the gate $U_{MQ}^{(12)}=\exp[i\pi(1-\eta_1^{(1)}\eta_1^{(2)})/4]$.

\section{Inner-Outer Majorana Basis}
Instead of forming Dirac fermions in the same TSC, one can instead form a full fermion from the MFs closest together (inner fermion) and a fermion from the MFs furthest apart (outer fermion),
\begin{align}
g_r&=(\gamma_{r}'+i\gamma_{l})/2\,,\nonumber\\
g_l&=(\gamma_{l}'+i\gamma_{r})/2\,,
\end{align}
respectively. The MFs are, in turn, written as
\begin{align}
\gamma_{\nu}'&=g_\nu+g_\nu^\dagger\,,\nonumber\\
\gamma_{\nu}&=(g_{\bar\nu}-g_{\bar\nu}^\dagger)/i\,.
\end{align} 
The tunneling Hamiltonian can then be written as
\begin{align}
\tilde H_T&=\sum_{\sigma,\nu}it_{\nu}d_\sigma^\dagger(g_\nu+g_\nu^\dagger)-it_{\nu}'d_\sigma^\dagger(g_{\bar\nu}-g_{\bar\nu}^\dagger)-i t_{\nu}'^*(g_{\bar\nu}-g_{\bar\nu}^\dagger)d_\sigma-it_{\nu}^*(g_\nu+g_\nu^\dagger)d_\sigma\nonumber\\
&=\sum_{\sigma,\nu}i d_\sigma^\dagger[(t_{\nu}-t_{R\bar\nu}) g_\nu+(t_{\nu}+t'_{\bar\nu})g_{\nu}^\dagger]-i[(t_{\nu}^*+t_{\bar\nu}'^*)g_\nu+(t_{\nu}^*-t_{\bar\nu}'^*)g_\nu^\dagger]d_\sigma\nonumber\\
&=\sum_{\sigma,\nu}-i \tilde t_{\nu -}d_\sigma^\dagger g_\nu+i \tilde t_{\nu -}^* g_{\nu}^\dagger d_\sigma + i \tilde t_{\nu +} d_\sigma^\dagger g_\nu^\dagger-i \tilde t_{\nu +}^* g_\nu d_\sigma\,,
\end{align}
where we have defined $\tilde t_{\nu -}=t'_{\bar\nu}-t_{\nu}$ and $\tilde t_{\nu +}=t_{\nu}+t'_{\bar\nu}$. Furthermore, we redefine the MF coupling in the TSC so that $\tilde H_M=\sum_\nu\tilde\delta_\nu(2g_\nu^\dagger g_\nu-1)$ where $\tilde\delta_r$ ($\tilde\delta_l$) now parameterizes the overlap between the inner (outer) MFs. With this redefinition, we see that the transformed Hamiltonian is, term by term, identical to Eq.~(\ref{Hsw}). Upon performing the same Schrieffer-Wolff transformation we find
\begin{align}
\tilde{\mathcal{H}}_T&=\tilde{\mathcal{H}}_s+\tilde{\mathcal H}_e+\tilde{\mathcal H_o}\,,\nonumber\\
\tilde{\mathcal H}_s&=\sum_{\sigma,\nu}\left(\frac{|\tilde t_{\nu +}|^2}{\epsilon_\sigma+2\tilde\delta_{\nu}}g_\nu^\dagger g_\nu +\frac{|\tilde t_{\nu -}|^2}{\epsilon_\sigma-2\tilde\delta_{\nu}}g_\nu g_\nu^\dagger\right)\left(2n_\sigma+d^\dagger_{\bar\sigma}d_\sigma+d^\dagger_{\sigma}d_{\bar\sigma}\right)\,,\nonumber\\
\tilde{\mathcal{H}}_o&=\sum_{\sigma,\nu}\left(\frac{n_\sigma+d^\dagger_{\bar\sigma}d_\sigma}{\epsilon_\sigma-2\tilde\delta_{\bar\nu}}+\frac{n_\sigma+d_\sigma^\dagger d_{\bar\sigma}}{{\epsilon_\sigma-2\tilde\delta_{\nu}}}\right)\tilde t^*_{\bar\nu -}\tilde t_{\nu -}g_\nu g_{\bar\nu}^\dagger+\left(\frac{n_\sigma+d^\dagger_{\bar\sigma}d_\sigma}{\epsilon_\sigma+2\tilde\delta_{\bar\nu}}+\frac{n_\sigma+d_\sigma^\dagger d_{\bar\sigma}}{\epsilon_{\sigma}+2\tilde\delta_{\nu}}\right)\tilde t^*_{\bar\nu +}\tilde t_{\nu +} g_\nu^\dagger g_{\bar\nu}\,,\nonumber\\
\tilde{\mathcal{H}}_e&=-\sum_{\sigma,\nu}\tilde t_{\bar\nu +}^*\tilde t_{\nu -}\left(\frac{n_\sigma+d_{\bar\sigma}^\dagger d_\sigma}{\epsilon_\sigma+2\tilde\delta_{\bar\nu}}+\frac{n_\sigma+d_{\sigma}^\dagger d_{\bar\sigma}}{\epsilon_{\sigma}-2\tilde\delta_{\nu}}\right)  g_\nu g_{\bar\nu} + \tilde t_{\bar\nu -}^*\tilde t_{\nu +}\left(\frac{n_\sigma+d_{\bar\sigma}^\dagger d_\sigma}{\epsilon_\sigma-2\tilde\delta_{\bar\nu}}+\frac{n_\sigma+d_{\sigma}^\dagger d_{\bar\sigma}}{\epsilon_{\sigma}+2\tilde\delta_{\nu}}\right)  g_\nu^\dagger g_{\bar\nu}^\dagger\,.
\end{align}
When the outer MFs are totally uncoupled to the system, $t_{r}=t_{l}'=0$, then $t_{ l\pm}=0$ so that
\begin{align}
\left(\frac{|\tilde t_{r+}|^2}{\epsilon_\sigma+2\tilde\delta_{r}}g_r^\dagger g_r +\frac{|\tilde t_{r-}|^2}{\epsilon_\sigma-2\tilde\delta_{r}}g_r g_r^\dagger\right)\left(2n_\sigma+d^\dagger_{\bar\sigma}d_\sigma+d^\dagger_{\sigma}d_{\bar\sigma}\right)\,.
\end{align}
One can immediately see that the effective magnetic field is, in general, different when the state is occupied versus unoccupied.

\section{Parity Measurement}
One can prepare the system so the initial state of the spin qubit is spin up and the MF qubit is in a superposition of eigenvalues of $\eta_1$, $|i\rangle=\left |\uparrow\right\rangle(\alpha|+\rangle+\beta|-\rangle)$ where $\eta_1|\pm\rangle=\pm|\pm\rangle= (|r\rangle\pm |l\rangle)/\sqrt{2}$. Rewriting the effective exchange Hamiltonian in terms of projection operators, $\mathcal P_{\pm}=(1\pm\eta_1)/2$, $(\mathbb 1+\sigma_1)\left[(|t_r|^2+|t_l|^2)+2\textrm{Re} t_r t_l^* \left(\mathcal P_+-\mathcal P_-\right)\right]/\epsilon_0$.
The time evolved initial state is
\begin{equation}
|i(\tau)\rangle
=(\alpha e^{i\omega_+ \tau}\cos(\omega_+ \tau)|+\rangle + \beta e^{i\omega_- \tau}\cos(\omega_- \tau)|-\rangle)\left |\uparrow\right\rangle+i(\alpha e^{i\omega_+ \tau}\sin(\omega_+ \tau)|+\rangle + \beta e^{i\omega_- \tau}\sin(\omega_- \tau)|-\rangle)|\downarrow\rangle\,,
\end{equation}
where $\omega_\pm=|t_r\pm t_l|^2$. In the simplest case when $t_r=t_l=t$, the probability to find the spin in the down state is $\mathbb P( |\downarrow\rangle) = |\alpha|^2\sin^2(4|t|^2 \tau/\hbar)$ and the probability to find the spin in the up state is $\mathbb P( \left |\uparrow\right\rangle) = 1-\mathbb P( \left |\downarrow\right\rangle)$. Coupling the spin and MF qubit for a time $\pi\hbar/8|t|^2$, the probability for the quantum dot to be in a spin up (down) state is equal to probability of finding the initial system in the $|-\rangle$ ($|+\rangle$) state, from which one can deduce the superposition of MF parity states.

One can alternatively use a basis of Dirac fermions which are formed from the inner [$g_r=(\gamma'_{r}+i\gamma_{l})/2$] and outer [$g_l=(\gamma'_{l}+i\gamma_{u})/2$] MFs of opposite TSCs. When the MFs on the same TSC are well separated, $\delta=t_{l}'=t_{r}=0$, the tunneling Hamiltonian in the new basis is $2\left(|t'_{+}|^2g_r^\dagger g_r +|t'_{-}|^2g_r g_r^\dagger\right)\left(\mathbb1+\sigma_1\right)/\epsilon_0$ where $t_\pm'=t_{r}'\pm t_{l}$. When the parity of the junction between the TSCs is one (zero), i.e. the complex fermion state formed by the inner MFs is occupied (unoccupied), there is an effective magnetic field on the dot proportional to $|t'_{-}|^2$ ($|t'_{+}|^2$). The Rabi oscillations between the spin up and down eigenstates, which can be detected, are therefore sensitive to the parity of the junction between two TSCs. The parity can be measured because the MF qubit is in a fixed parity subspace, i.e. if the fermion is not shared by the nearest MFs then the it must be shared between the outer MFs. If the parity is unrestricted, one must measure both MFs on both the left and right TSCs to determine the state of the MF qubit. 

\end{widetext}

\begin{thebibliography}{48}%
\makeatletter
\providecommand \@ifxundefined [1]{%
 \@ifx{#1\undefined}
}%
\providecommand \@ifnum [1]{%
 \ifnum #1\expandafter \@firstoftwo
 \else \expandafter \@secondoftwo
 \fi
}%
\providecommand \@ifx [1]{%
 \ifx #1\expandafter \@firstoftwo
 \else \expandafter \@secondoftwo
 \fi
}%
\providecommand \natexlab [1]{#1}%
\providecommand \enquote  [1]{``#1''}%
\providecommand \bibnamefont  [1]{#1}%
\providecommand \bibfnamefont [1]{#1}%
\providecommand \citenamefont [1]{#1}%
\providecommand \href@noop [0]{\@secondoftwo}%
\providecommand \href [0]{\begingroup \@sanitize@url \@href}%
\providecommand \@href[1]{\@@startlink{#1}\@@href}%
\providecommand \@@href[1]{\endgroup#1\@@endlink}%
\providecommand \@sanitize@url [0]{\catcode `\\12\catcode `\$12\catcode
  `\&12\catcode `\#12\catcode `\^12\catcode `\_12\catcode `\%12\relax}%
\providecommand \@@startlink[1]{}%
\providecommand \@@endlink[0]{}%
\providecommand \url  [0]{\begingroup\@sanitize@url \@url }%
\providecommand \@url [1]{\endgroup\@href {#1}{\urlprefix }}%
\providecommand \urlprefix  [0]{URL }%
\providecommand \Eprint [0]{\href }%
\providecommand \doibase [0]{http://dx.doi.org/}%
\providecommand \selectlanguage [0]{\@gobble}%
\providecommand \bibinfo  [0]{\@secondoftwo}%
\providecommand \bibfield  [0]{\@secondoftwo}%
\providecommand \translation [1]{[#1]}%
\providecommand \BibitemOpen [0]{}%
\providecommand \bibitemStop [0]{}%
\providecommand \bibitemNoStop [0]{.\EOS\space}%
\providecommand \EOS [0]{\spacefactor3000\relax}%
\providecommand \BibitemShut  [1]{\csname bibitem#1\endcsname}%
\let\auto@bib@innerbib\@empty
%</preamble>
\bibitem [{\citenamefont {Loss}\ and\ \citenamefont
  {DiVincenzo}(1998)}]{lossPRA98}%
  \BibitemOpen
  \bibfield  {author} {\bibinfo {author} {\bibfnamefont {D.}~\bibnamefont
  {Loss}}\ and\ \bibinfo {author} {\bibfnamefont {D.~P.}\ \bibnamefont
  {DiVincenzo}},\ }\href@noop {} {\bibfield  {journal} {\bibinfo  {journal}
  {Phys. Rev. A}\ }\textbf {\bibinfo {volume} {57}},\ \bibinfo {pages} {120}
  (\bibinfo {year} {1998})}\BibitemShut {NoStop}%
\bibitem [{\citenamefont {Hanson}\ \emph {et~al.}(2007)\citenamefont {Hanson},
  \citenamefont {Kouwenhoven}, \citenamefont {Petta}, \citenamefont {Tarucha},\
  and\ \citenamefont {Vandersypen}}]{hansonRMP07}%
  \BibitemOpen
  \bibfield  {author} {\bibinfo {author} {\bibfnamefont {R.}~\bibnamefont
  {Hanson}}, \bibinfo {author} {\bibfnamefont {L.~P.}\ \bibnamefont
  {Kouwenhoven}}, \bibinfo {author} {\bibfnamefont {J.~R.}\ \bibnamefont
  {Petta}}, \bibinfo {author} {\bibfnamefont {S.}~\bibnamefont {Tarucha}}, \
  and\ \bibinfo {author} {\bibfnamefont {L.~M.~K.}\ \bibnamefont
  {Vandersypen}},\ }\href {\doibase 10.1103/RevModPhys.79.1217} {\bibfield
  {journal} {\bibinfo  {journal} {Rev. Mod. Phys.}\ }\textbf {\bibinfo {volume}
  {79}},\ \bibinfo {pages} {1217} (\bibinfo {year} {2007})}\BibitemShut
  {NoStop}%
\bibitem [{\citenamefont {Kloeffel}\ and\ \citenamefont
  {Loss}(2013)}]{kloeffelARCMP13}%
  \BibitemOpen
  \bibfield  {author} {\bibinfo {author} {\bibfnamefont {C.}~\bibnamefont
  {Kloeffel}}\ and\ \bibinfo {author} {\bibfnamefont {D.}~\bibnamefont
  {Loss}},\ }\href@noop {} {\bibfield  {journal} {\bibinfo  {journal} {Annu.
  Rev. Condens. Matter Phys.}\ }\textbf {\bibinfo {volume} {4}},\ \bibinfo
  {pages} {51} (\bibinfo {year} {2013})}\BibitemShut {NoStop}%
\bibitem [{\citenamefont {Nayak}\ \emph {et~al.}(2008)\citenamefont {Nayak},
  \citenamefont {Simon}, \citenamefont {Stern}, \citenamefont {Freedman},\ and\
  \citenamefont {Das~Sarma}}]{nayakRMP08}%
  \BibitemOpen
  \bibfield  {author} {\bibinfo {author} {\bibfnamefont {C.}~\bibnamefont
  {Nayak}}, \bibinfo {author} {\bibfnamefont {S.~H.}\ \bibnamefont {Simon}},
  \bibinfo {author} {\bibfnamefont {A.}~\bibnamefont {Stern}}, \bibinfo
  {author} {\bibfnamefont {M.}~\bibnamefont {Freedman}}, \ and\ \bibinfo
  {author} {\bibfnamefont {S.}~\bibnamefont {Das~Sarma}},\ }\href {\doibase
  10.1103/RevModPhys.80.1083} {\bibfield  {journal} {\bibinfo  {journal} {Rev.
  Mod. Phys.}\ }\textbf {\bibinfo {volume} {80}},\ \bibinfo {pages} {1083}
  (\bibinfo {year} {2008})}\BibitemShut {NoStop}%
\bibitem [{\citenamefont {Goldin}\ \emph {et~al.}(1985)\citenamefont {Goldin},
  \citenamefont {Menikoff},\ and\ \citenamefont {Sharp}}]{goldinPRL85}%
  \BibitemOpen
  \bibfield  {author} {\bibinfo {author} {\bibfnamefont {G.~A.}\ \bibnamefont
  {Goldin}}, \bibinfo {author} {\bibfnamefont {R.}~\bibnamefont {Menikoff}}, \
  and\ \bibinfo {author} {\bibfnamefont {D.~H.}\ \bibnamefont {Sharp}},\
  }\href@noop {} {\bibfield  {journal} {\bibinfo  {journal} {Phys. Rev. Lett.}\
  }\textbf {\bibinfo {volume} {54}},\ \bibinfo {pages} {603} (\bibinfo {year}
  {1985})}\BibitemShut {NoStop}%
\bibitem [{\citenamefont {Kitaev}(2003)}]{kitaevAoP03}%
  \BibitemOpen
  \bibfield  {author} {\bibinfo {author} {\bibfnamefont {A.~Y.}\ \bibnamefont
  {Kitaev}},\ }\href@noop {} {\bibfield  {journal} {\bibinfo  {journal} {Ann.
  Phys.}\ }\textbf {\bibinfo {volume} {303}},\ \bibinfo {pages} {2} (\bibinfo
  {year} {2003})}\BibitemShut {NoStop}%
\bibitem [{\citenamefont {Freedman}\ \emph {et~al.}(2002)\citenamefont
  {Freedman}, \citenamefont {Kitaev},\ and\ \citenamefont
  {Wang}}]{freedmanCiMP02}%
  \BibitemOpen
  \bibfield  {author} {\bibinfo {author} {\bibfnamefont {H.~M.}\ \bibnamefont
  {Freedman}}, \bibinfo {author} {\bibfnamefont {A.}~\bibnamefont {Kitaev}}, \
  and\ \bibinfo {author} {\bibfnamefont {Z.}~\bibnamefont {Wang}},\ }\href
  {\doibase 10.1007/s002200200635} {\bibfield  {journal} {\bibinfo  {journal}
  {Commun. Math. Phys.}\ }\textbf {\bibinfo {volume} {227}},\ \bibinfo {pages}
  {587} (\bibinfo {year} {2002})}\BibitemShut {NoStop}%
\bibitem [{\citenamefont {Bravyi}\ and\ \citenamefont
  {Kitaev}(2002)}]{bravyiAoP02}%
  \BibitemOpen
  \bibfield  {author} {\bibinfo {author} {\bibfnamefont {S.~B.}\ \bibnamefont
  {Bravyi}}\ and\ \bibinfo {author} {\bibfnamefont {A.~Y.}\ \bibnamefont
  {Kitaev}},\ }\href {\doibase http://dx.doi.org/10.1006/aphy.2002.6254}
  {\bibfield  {journal} {\bibinfo  {journal} {Annals of Physics}\ }\textbf
  {\bibinfo {volume} {298}},\ \bibinfo {pages} {210 } (\bibinfo {year}
  {2002})}\BibitemShut {NoStop}%
\bibitem [{\citenamefont {Freedman}\ \emph {et~al.}(2003)\citenamefont
  {Freedman}, \citenamefont {Kitaev}, \citenamefont {Larsen},\ and\
  \citenamefont {Wang}}]{freedmanBAMS03}%
  \BibitemOpen
  \bibfield  {author} {\bibinfo {author} {\bibfnamefont {M.~H.}\ \bibnamefont
  {Freedman}}, \bibinfo {author} {\bibfnamefont {A.}~\bibnamefont {Kitaev}},
  \bibinfo {author} {\bibfnamefont {M.~J.}\ \bibnamefont {Larsen}}, \ and\
  \bibinfo {author} {\bibfnamefont {Z.}~\bibnamefont {Wang}},\ }\href {\doibase
  10.1090/S0273-0979-02-00964-3} {\bibfield  {journal} {\bibinfo  {journal}
  {Bull. Amer. Math. Soc. (N.S.)}\ }\textbf {\bibinfo {volume} {40}},\ \bibinfo
  {pages} {31} (\bibinfo {year} {2003})}\BibitemShut {NoStop}%
\bibitem [{\citenamefont {Mourik}\ \emph {et~al.}(2012)\citenamefont {Mourik},
  \citenamefont {Zuo}, \citenamefont {Frolov}, \citenamefont {Plissard},
  \citenamefont {Bakkers},\ and\ \citenamefont {Kouwenhoven}}]{mourikSCI12}%
  \BibitemOpen
  \bibfield  {author} {\bibinfo {author} {\bibfnamefont {V.}~\bibnamefont
  {Mourik}}, \bibinfo {author} {\bibfnamefont {K.}~\bibnamefont {Zuo}},
  \bibinfo {author} {\bibfnamefont {S.}~\bibnamefont {Frolov}}, \bibinfo
  {author} {\bibfnamefont {S.}~\bibnamefont {Plissard}}, \bibinfo {author}
  {\bibfnamefont {E.}~\bibnamefont {Bakkers}}, \ and\ \bibinfo {author}
  {\bibfnamefont {L.}~\bibnamefont {Kouwenhoven}},\ }\href@noop {} {\bibfield
  {journal} {\bibinfo  {journal} {Science}\ }\textbf {\bibinfo {volume}
  {336}},\ \bibinfo {pages} {1003} (\bibinfo {year} {2012})}\BibitemShut
  {NoStop}%
\bibitem [{\citenamefont {Das}\ \emph {et~al.}(2012)\citenamefont {Das},
  \citenamefont {Ronen}, \citenamefont {Most}, \citenamefont {Oreg},
  \citenamefont {Heiblum},\ and\ \citenamefont {Shtrikman}}]{dasNATP12}%
  \BibitemOpen
  \bibfield  {author} {\bibinfo {author} {\bibfnamefont {A.}~\bibnamefont
  {Das}}, \bibinfo {author} {\bibfnamefont {Y.}~\bibnamefont {Ronen}}, \bibinfo
  {author} {\bibfnamefont {Y.}~\bibnamefont {Most}}, \bibinfo {author}
  {\bibfnamefont {Y.}~\bibnamefont {Oreg}}, \bibinfo {author} {\bibfnamefont
  {M.}~\bibnamefont {Heiblum}}, \ and\ \bibinfo {author} {\bibfnamefont
  {H.}~\bibnamefont {Shtrikman}},\ }\href@noop {} {\bibfield  {journal}
  {\bibinfo  {journal} {Nat. Phys.}\ }\textbf {\bibinfo {volume} {8}},\
  \bibinfo {pages} {887} (\bibinfo {year} {2012})}\BibitemShut {NoStop}%
\bibitem [{\citenamefont {Rokhinson}\ \emph {et~al.}(2012)\citenamefont
  {Rokhinson}, \citenamefont {Liu},\ and\ \citenamefont
  {Furdyna}}]{rokhinsonNATP12}%
  \BibitemOpen
  \bibfield  {author} {\bibinfo {author} {\bibfnamefont {L.~P.}\ \bibnamefont
  {Rokhinson}}, \bibinfo {author} {\bibfnamefont {X.}~\bibnamefont {Liu}}, \
  and\ \bibinfo {author} {\bibfnamefont {J.~K.}\ \bibnamefont {Furdyna}},\
  }\href {http://dx.doi.org/10.1038/nphys2429} {\bibfield  {journal} {\bibinfo
  {journal} {Nat. Phys.}\ }\textbf {\bibinfo {volume} {8}},\ \bibinfo {pages}
  {795} (\bibinfo {year} {2012})}\BibitemShut {NoStop}%
\bibitem [{\citenamefont {Deng}\ \emph {et~al.}(2012)\citenamefont {Deng},
  \citenamefont {Yu}, \citenamefont {Huang}, \citenamefont {Larsson},
  \citenamefont {Caroff},\ and\ \citenamefont {Xu}}]{dengNANOL12}%
  \BibitemOpen
  \bibfield  {author} {\bibinfo {author} {\bibfnamefont {M.~T.}\ \bibnamefont
  {Deng}}, \bibinfo {author} {\bibfnamefont {C.~L.}\ \bibnamefont {Yu}},
  \bibinfo {author} {\bibfnamefont {G.~Y.}\ \bibnamefont {Huang}}, \bibinfo
  {author} {\bibfnamefont {M.}~\bibnamefont {Larsson}}, \bibinfo {author}
  {\bibfnamefont {P.}~\bibnamefont {Caroff}}, \ and\ \bibinfo {author}
  {\bibfnamefont {H.~Q.}\ \bibnamefont {Xu}},\ }\href {\doibase
  10.1021/nl303758w} {\bibfield  {journal} {\bibinfo  {journal} {Nano Letters}\
  }\textbf {\bibinfo {volume} {12}},\ \bibinfo {pages} {6414} (\bibinfo {year}
  {2012})}\BibitemShut {NoStop}%
\bibitem [{\citenamefont {Finck}\ \emph {et~al.}(2013)\citenamefont {Finck},
  \citenamefont {Van~Harlingen}, \citenamefont {Mohseni}, \citenamefont
  {Jung},\ and\ \citenamefont {Li}}]{finckPRL13}%
  \BibitemOpen
  \bibfield  {author} {\bibinfo {author} {\bibfnamefont {A.~D.~K.}\
  \bibnamefont {Finck}}, \bibinfo {author} {\bibfnamefont {D.~J.}\ \bibnamefont
  {Van~Harlingen}}, \bibinfo {author} {\bibfnamefont {P.~K.}\ \bibnamefont
  {Mohseni}}, \bibinfo {author} {\bibfnamefont {K.}~\bibnamefont {Jung}}, \
  and\ \bibinfo {author} {\bibfnamefont {X.}~\bibnamefont {Li}},\ }\href
  {\doibase 10.1103/PhysRevLett.110.126406} {\bibfield  {journal} {\bibinfo
  {journal} {Phys. Rev. Lett.}\ }\textbf {\bibinfo {volume} {110}},\ \bibinfo
  {pages} {126406} (\bibinfo {year} {2013})}\BibitemShut {NoStop}%
\bibitem [{\citenamefont {Churchill}\ \emph {et~al.}(2013)\citenamefont
  {Churchill}, \citenamefont {Fatemi}, \citenamefont {Grove-Rasmussen},
  \citenamefont {Deng}, \citenamefont {Caroff}, \citenamefont {Xu},\ and\
  \citenamefont {Marcus}}]{churchillPRB13}%
  \BibitemOpen
  \bibfield  {author} {\bibinfo {author} {\bibfnamefont {H.~O.~H.}\
  \bibnamefont {Churchill}}, \bibinfo {author} {\bibfnamefont {V.}~\bibnamefont
  {Fatemi}}, \bibinfo {author} {\bibfnamefont {K.}~\bibnamefont
  {Grove-Rasmussen}}, \bibinfo {author} {\bibfnamefont {M.~T.}\ \bibnamefont
  {Deng}}, \bibinfo {author} {\bibfnamefont {P.}~\bibnamefont {Caroff}},
  \bibinfo {author} {\bibfnamefont {H.~Q.}\ \bibnamefont {Xu}}, \ and\ \bibinfo
  {author} {\bibfnamefont {C.~M.}\ \bibnamefont {Marcus}},\ }\href {\doibase
  10.1103/PhysRevB.87.241401} {\bibfield  {journal} {\bibinfo  {journal} {Phys.
  Rev. B}\ }\textbf {\bibinfo {volume} {87}},\ \bibinfo {pages} {241401}
  (\bibinfo {year} {2013})}\BibitemShut {NoStop}%
\bibitem [{\citenamefont {Nadj-Perge}\ \emph {et~al.}(2014)\citenamefont
  {Nadj-Perge}, \citenamefont {Drozdov}, \citenamefont {Li}, \citenamefont
  {Chen}, \citenamefont {Jeon}, \citenamefont {Seo}, \citenamefont {MacDonald},
  \citenamefont {Bernevig},\ and\ \citenamefont {Yazdani}}]{nadj-pergeSCI14}%
  \BibitemOpen
  \bibfield  {author} {\bibinfo {author} {\bibfnamefont {S.}~\bibnamefont
  {Nadj-Perge}}, \bibinfo {author} {\bibfnamefont {I.~K.}\ \bibnamefont
  {Drozdov}}, \bibinfo {author} {\bibfnamefont {J.}~\bibnamefont {Li}},
  \bibinfo {author} {\bibfnamefont {H.}~\bibnamefont {Chen}}, \bibinfo {author}
  {\bibfnamefont {S.}~\bibnamefont {Jeon}}, \bibinfo {author} {\bibfnamefont
  {J.}~\bibnamefont {Seo}}, \bibinfo {author} {\bibfnamefont {A.~H.}\
  \bibnamefont {MacDonald}}, \bibinfo {author} {\bibfnamefont {B.~A.}\
  \bibnamefont {Bernevig}}, \ and\ \bibinfo {author} {\bibfnamefont
  {A.}~\bibnamefont {Yazdani}},\ }\href {\doibase 10.1126/science.1259327}
  {\bibfield  {journal} {\bibinfo  {journal} {Science}\ }\textbf {\bibinfo
  {volume} {346}},\ \bibinfo {pages} {602} (\bibinfo {year}
  {2014})}\BibitemShut {NoStop}%
\bibitem [{\citenamefont {Pawlak}\ \emph {et~al.}(2015)\citenamefont {Pawlak},
  \citenamefont {Kisiel}, \citenamefont {Klinovaja}, \citenamefont {Meier},
  \citenamefont {Kawai}, \citenamefont {Glatzel}, \citenamefont {Loss},\ and\
  \citenamefont {Meyer}}]{pawlakCM15}%
  \BibitemOpen
  \bibfield  {author} {\bibinfo {author} {\bibfnamefont {R.}~\bibnamefont
  {Pawlak}}, \bibinfo {author} {\bibfnamefont {M.}~\bibnamefont {Kisiel}},
  \bibinfo {author} {\bibfnamefont {J.}~\bibnamefont {Klinovaja}}, \bibinfo
  {author} {\bibfnamefont {T.}~\bibnamefont {Meier}}, \bibinfo {author}
  {\bibfnamefont {S.}~\bibnamefont {Kawai}}, \bibinfo {author} {\bibfnamefont
  {T.}~\bibnamefont {Glatzel}}, \bibinfo {author} {\bibfnamefont
  {D.}~\bibnamefont {Loss}}, \ and\ \bibinfo {author} {\bibfnamefont
  {E.}~\bibnamefont {Meyer}},\ }\href@noop {} {\bibfield  {journal} {\bibinfo
  {journal} {arXiv preprint arXiv:1505.06078}\ } (\bibinfo {year}
  {2015})}\BibitemShut {NoStop}%
\bibitem [{\citenamefont {Lutchyn}\ \emph {et~al.}(2010)\citenamefont
  {Lutchyn}, \citenamefont {Sau},\ and\ \citenamefont {Sarma}}]{lutchynPRL10}%
  \BibitemOpen
  \bibfield  {author} {\bibinfo {author} {\bibfnamefont {R.~M.}\ \bibnamefont
  {Lutchyn}}, \bibinfo {author} {\bibfnamefont {J.~D.}\ \bibnamefont {Sau}}, \
  and\ \bibinfo {author} {\bibfnamefont {S.~D.}\ \bibnamefont {Sarma}},\
  }\href@noop {} {\bibfield  {journal} {\bibinfo  {journal} {Phys. Rev. Lett.}\
  }\textbf {\bibinfo {volume} {105}},\ \bibinfo {pages} {077001} (\bibinfo
  {year} {2010})}\BibitemShut {NoStop}%
\bibitem [{\citenamefont {Oreg}\ \emph {et~al.}(2010)\citenamefont {Oreg},
  \citenamefont {Refael},\ and\ \citenamefont {von Oppen}}]{oregPRL10}%
  \BibitemOpen
  \bibfield  {author} {\bibinfo {author} {\bibfnamefont {Y.}~\bibnamefont
  {Oreg}}, \bibinfo {author} {\bibfnamefont {G.}~\bibnamefont {Refael}}, \ and\
  \bibinfo {author} {\bibfnamefont {F.}~\bibnamefont {von Oppen}},\ }\href@noop
  {} {\bibfield  {journal} {\bibinfo  {journal} {Phys. Rev. Lett.}\ }\textbf
  {\bibinfo {volume} {105}},\ \bibinfo {pages} {177002} (\bibinfo {year}
  {2010})}\BibitemShut {NoStop}%
\bibitem [{\citenamefont {Fu}\ and\ \citenamefont {Kane}(2008)}]{fuPRL08}%
  \BibitemOpen
  \bibfield  {author} {\bibinfo {author} {\bibfnamefont {L.}~\bibnamefont
  {Fu}}\ and\ \bibinfo {author} {\bibfnamefont {C.~L.}\ \bibnamefont {Kane}},\
  }\href@noop {} {\bibfield  {journal} {\bibinfo  {journal} {Phys. Rev. Lett.}\
  }\textbf {\bibinfo {volume} {100}},\ \bibinfo {pages} {096407} (\bibinfo
  {year} {2008})}\BibitemShut {NoStop}%
\bibitem [{\citenamefont {Volovik}(1999)}]{volovikJETP99}%
  \BibitemOpen
  \bibfield  {author} {\bibinfo {author} {\bibfnamefont {G.}~\bibnamefont
  {Volovik}},\ }\href@noop {} {\bibfield  {journal} {\bibinfo  {journal} {JETP
  Letters}\ }\textbf {\bibinfo {volume} {70}},\ \bibinfo {pages} {609}
  (\bibinfo {year} {1999})}\BibitemShut {NoStop}%
\bibitem [{\citenamefont {Klinovaja}\ \emph {et~al.}(2013)\citenamefont
  {Klinovaja}, \citenamefont {Stano}, \citenamefont {Yazdani},\ and\
  \citenamefont {Loss}}]{klinovajaPRL13}%
  \BibitemOpen
  \bibfield  {author} {\bibinfo {author} {\bibfnamefont {J.}~\bibnamefont
  {Klinovaja}}, \bibinfo {author} {\bibfnamefont {P.}~\bibnamefont {Stano}},
  \bibinfo {author} {\bibfnamefont {A.}~\bibnamefont {Yazdani}}, \ and\
  \bibinfo {author} {\bibfnamefont {D.}~\bibnamefont {Loss}},\ }\href {\doibase
  10.1103/PhysRevLett.111.186805} {\bibfield  {journal} {\bibinfo  {journal}
  {Phys. Rev. Lett.}\ }\textbf {\bibinfo {volume} {111}},\ \bibinfo {pages}
  {186805} (\bibinfo {year} {2013})}\BibitemShut {NoStop}%
\bibitem [{\citenamefont {Vazifeh}\ and\ \citenamefont
  {Franz}(2013)}]{vazifehPRL13}%
  \BibitemOpen
  \bibfield  {author} {\bibinfo {author} {\bibfnamefont {M.~M.}\ \bibnamefont
  {Vazifeh}}\ and\ \bibinfo {author} {\bibfnamefont {M.}~\bibnamefont
  {Franz}},\ }\href {\doibase 10.1103/PhysRevLett.111.206802} {\bibfield
  {journal} {\bibinfo  {journal} {Phys. Rev. Lett.}\ }\textbf {\bibinfo
  {volume} {111}},\ \bibinfo {pages} {206802} (\bibinfo {year}
  {2013})}\BibitemShut {NoStop}%
\bibitem [{\citenamefont {Braunecker}\ and\ \citenamefont
  {Simon}(2013)}]{brauneckerPRL13}%
  \BibitemOpen
  \bibfield  {author} {\bibinfo {author} {\bibfnamefont {B.}~\bibnamefont
  {Braunecker}}\ and\ \bibinfo {author} {\bibfnamefont {P.}~\bibnamefont
  {Simon}},\ }\href {\doibase 10.1103/PhysRevLett.111.147202} {\bibfield
  {journal} {\bibinfo  {journal} {Phys. Rev. Lett.}\ }\textbf {\bibinfo
  {volume} {111}},\ \bibinfo {pages} {147202} (\bibinfo {year}
  {2013})}\BibitemShut {NoStop}%
\bibitem [{\citenamefont {Nadj-Perge}\ \emph {et~al.}(2013)\citenamefont
  {Nadj-Perge}, \citenamefont {Drozdov}, \citenamefont {Bernevig},\ and\
  \citenamefont {Yazdani}}]{nadj-pergePRB13}%
  \BibitemOpen
  \bibfield  {author} {\bibinfo {author} {\bibfnamefont {S.}~\bibnamefont
  {Nadj-Perge}}, \bibinfo {author} {\bibfnamefont {I.~K.}\ \bibnamefont
  {Drozdov}}, \bibinfo {author} {\bibfnamefont {B.~A.}\ \bibnamefont
  {Bernevig}}, \ and\ \bibinfo {author} {\bibfnamefont {A.}~\bibnamefont
  {Yazdani}},\ }\href {\doibase 10.1103/PhysRevB.88.020407} {\bibfield
  {journal} {\bibinfo  {journal} {Phys. Rev. B}\ }\textbf {\bibinfo {volume}
  {88}},\ \bibinfo {pages} {020407} (\bibinfo {year} {2013})}\BibitemShut
  {NoStop}%
\bibitem [{\citenamefont {Pientka}\ \emph {et~al.}(2013)\citenamefont
  {Pientka}, \citenamefont {Glazman},\ and\ \citenamefont {von
  Oppen}}]{pientkaPRB13}%
  \BibitemOpen
  \bibfield  {author} {\bibinfo {author} {\bibfnamefont {F.}~\bibnamefont
  {Pientka}}, \bibinfo {author} {\bibfnamefont {L.~I.}\ \bibnamefont
  {Glazman}}, \ and\ \bibinfo {author} {\bibfnamefont {F.}~\bibnamefont {von
  Oppen}},\ }\href {\doibase 10.1103/PhysRevB.88.155420} {\bibfield  {journal}
  {\bibinfo  {journal} {Phys. Rev. B}\ }\textbf {\bibinfo {volume} {88}},\
  \bibinfo {pages} {155420} (\bibinfo {year} {2013})}\BibitemShut {NoStop}%
\bibitem [{\citenamefont {Bravyi}(2006)}]{bravyiPRA06}%
  \BibitemOpen
  \bibfield  {author} {\bibinfo {author} {\bibfnamefont {S.}~\bibnamefont
  {Bravyi}},\ }\href {\doibase 10.1103/PhysRevA.73.042313} {\bibfield
  {journal} {\bibinfo  {journal} {Phys. Rev. A}\ }\textbf {\bibinfo {volume}
  {73}},\ \bibinfo {pages} {042313} (\bibinfo {year} {2006})}\BibitemShut
  {NoStop}%
\bibitem [{\citenamefont {Hyart}\ \emph {et~al.}(2013)\citenamefont {Hyart},
  \citenamefont {van Heck}, \citenamefont {Fulga}, \citenamefont {Burrello},
  \citenamefont {Akhmerov},\ and\ \citenamefont {Beenakker}}]{hyartPRB13}%
  \BibitemOpen
  \bibfield  {author} {\bibinfo {author} {\bibfnamefont {T.}~\bibnamefont
  {Hyart}}, \bibinfo {author} {\bibfnamefont {B.}~\bibnamefont {van Heck}},
  \bibinfo {author} {\bibfnamefont {I.~C.}\ \bibnamefont {Fulga}}, \bibinfo
  {author} {\bibfnamefont {M.}~\bibnamefont {Burrello}}, \bibinfo {author}
  {\bibfnamefont {A.~R.}\ \bibnamefont {Akhmerov}}, \ and\ \bibinfo {author}
  {\bibfnamefont {C.~W.~J.}\ \bibnamefont {Beenakker}},\ }\href {\doibase
  10.1103/PhysRevB.88.035121} {\bibfield  {journal} {\bibinfo  {journal} {Phys.
  Rev. B}\ }\textbf {\bibinfo {volume} {88}},\ \bibinfo {pages} {035121}
  (\bibinfo {year} {2013})}\BibitemShut {NoStop}%
\bibitem [{\citenamefont {Leijnse}\ and\ \citenamefont
  {Flensberg}(2011)}]{leijnsePRL11}%
  \BibitemOpen
  \bibfield  {author} {\bibinfo {author} {\bibfnamefont {M.}~\bibnamefont
  {Leijnse}}\ and\ \bibinfo {author} {\bibfnamefont {K.}~\bibnamefont
  {Flensberg}},\ }\href@noop {} {\bibfield  {journal} {\bibinfo  {journal}
  {Phys. Rev. Lett.}\ }\textbf {\bibinfo {volume} {107}},\ \bibinfo {pages}
  {210502} (\bibinfo {year} {2011})}\BibitemShut {NoStop}%
\bibitem [{\citenamefont {Kitaev}(2001)}]{kitaevPU01}%
  \BibitemOpen
  \bibfield  {author} {\bibinfo {author} {\bibfnamefont {A.~Y.}\ \bibnamefont
  {Kitaev}},\ }\href@noop {} {\bibfield  {journal} {\bibinfo  {journal}
  {Physics-Uspekhi}\ }\textbf {\bibinfo {volume} {44}},\ \bibinfo {pages} {131}
  (\bibinfo {year} {2001})}\BibitemShut {NoStop}%
\bibitem [{\citenamefont {Rainis}\ \emph {et~al.}(2013)\citenamefont {Rainis},
  \citenamefont {Trifunovic}, \citenamefont {Klinovaja},\ and\ \citenamefont
  {Loss}}]{rainisPRB13}%
  \BibitemOpen
  \bibfield  {author} {\bibinfo {author} {\bibfnamefont {D.}~\bibnamefont
  {Rainis}}, \bibinfo {author} {\bibfnamefont {L.}~\bibnamefont {Trifunovic}},
  \bibinfo {author} {\bibfnamefont {J.}~\bibnamefont {Klinovaja}}, \ and\
  \bibinfo {author} {\bibfnamefont {D.}~\bibnamefont {Loss}},\ }\href {\doibase
  10.1103/PhysRevB.87.024515} {\bibfield  {journal} {\bibinfo  {journal} {Phys.
  Rev. B}\ }\textbf {\bibinfo {volume} {87}},\ \bibinfo {pages} {024515}
  (\bibinfo {year} {2013})}\BibitemShut {NoStop}%
\bibitem [{\citenamefont {Zyuzin}\ \emph {et~al.}(2013)\citenamefont {Zyuzin},
  \citenamefont {Rainis}, \citenamefont {Klinovaja},\ and\ \citenamefont
  {Loss}}]{zyuzinPRL13}%
  \BibitemOpen
  \bibfield  {author} {\bibinfo {author} {\bibfnamefont {A.}~\bibnamefont
  {Zyuzin}}, \bibinfo {author} {\bibfnamefont {D.}~\bibnamefont {Rainis}},
  \bibinfo {author} {\bibfnamefont {J.}~\bibnamefont {Klinovaja}}, \ and\
  \bibinfo {author} {\bibfnamefont {D.}~\bibnamefont {Loss}},\ }\href@noop {}
  {\bibfield  {journal} {\bibinfo  {journal} {Phys. Rev. Lett.}\ }\textbf
  {\bibinfo {volume} {111}},\ \bibinfo {pages} {056802} (\bibinfo {year}
  {2013})}\BibitemShut {NoStop}%
\bibitem [{\citenamefont {Rainis}\ and\ \citenamefont
  {Loss}(2012)}]{rainisPRB12}%
  \BibitemOpen
  \bibfield  {author} {\bibinfo {author} {\bibfnamefont {D.}~\bibnamefont
  {Rainis}}\ and\ \bibinfo {author} {\bibfnamefont {D.}~\bibnamefont {Loss}},\
  }\href@noop {} {\bibfield  {journal} {\bibinfo  {journal} {Phys. Rev. B}\
  }\textbf {\bibinfo {volume} {85}},\ \bibinfo {pages} {174533} (\bibinfo
  {year} {2012})}\BibitemShut {NoStop}%
\bibitem [{\citenamefont {Pedrocchi}\ and\ \citenamefont
  {DiVincenzo}(2015)}]{pedrocchiPRL15}%
  \BibitemOpen
  \bibfield  {author} {\bibinfo {author} {\bibfnamefont {F.~L.}\ \bibnamefont
  {Pedrocchi}}\ and\ \bibinfo {author} {\bibfnamefont {D.~P.}\ \bibnamefont
  {DiVincenzo}},\ }\href {\doibase 10.1103/PhysRevLett.115.120402} {\bibfield
  {journal} {\bibinfo  {journal} {Phys. Rev. Lett.}\ }\textbf {\bibinfo
  {volume} {115}},\ \bibinfo {pages} {120402} (\bibinfo {year}
  {2015})}\BibitemShut {NoStop}%
\bibitem [{\citenamefont {Hutter}\ and\ \citenamefont
  {Wootton}(2015)}]{hutterCM15}%
  \BibitemOpen
  \bibfield  {author} {\bibinfo {author} {\bibfnamefont {A.}~\bibnamefont
  {Hutter}}\ and\ \bibinfo {author} {\bibfnamefont {J.~R.}\ \bibnamefont
  {Wootton}},\ }\href@noop {} {\bibfield  {journal} {\bibinfo  {journal} {arXiv
  preprint arXiv:1508.04033}\ } (\bibinfo {year} {2015})}\BibitemShut {NoStop}%
\bibitem [{\citenamefont {Tewari}\ \emph {et~al.}(2008)\citenamefont {Tewari},
  \citenamefont {Zhang}, \citenamefont {Sarma}, \citenamefont {Nayak},\ and\
  \citenamefont {Lee}}]{tewariPRL08}%
  \BibitemOpen
  \bibfield  {author} {\bibinfo {author} {\bibfnamefont {S.}~\bibnamefont
  {Tewari}}, \bibinfo {author} {\bibfnamefont {C.}~\bibnamefont {Zhang}},
  \bibinfo {author} {\bibfnamefont {S.~D.}\ \bibnamefont {Sarma}}, \bibinfo
  {author} {\bibfnamefont {C.}~\bibnamefont {Nayak}}, \ and\ \bibinfo {author}
  {\bibfnamefont {D.-H.}\ \bibnamefont {Lee}},\ }\href@noop {} {\bibfield
  {journal} {\bibinfo  {journal} {Phys. Rev. Lett.}\ }\textbf {\bibinfo
  {volume} {100}},\ \bibinfo {pages} {027001} (\bibinfo {year}
  {2008})}\BibitemShut {NoStop}%
\bibitem [{\citenamefont {Braun}\ \emph {et~al.}(2004)\citenamefont {Braun},
  \citenamefont {K\"onig},\ and\ \citenamefont {Martinek}}]{braunPRB04}%
  \BibitemOpen
  \bibfield  {author} {\bibinfo {author} {\bibfnamefont {M.}~\bibnamefont
  {Braun}}, \bibinfo {author} {\bibfnamefont {J.}~\bibnamefont {K\"onig}}, \
  and\ \bibinfo {author} {\bibfnamefont {J.}~\bibnamefont {Martinek}},\ }\href
  {\doibase 10.1103/PhysRevB.70.195345} {\bibfield  {journal} {\bibinfo
  {journal} {Phys. Rev. B}\ }\textbf {\bibinfo {volume} {70}},\ \bibinfo
  {pages} {195345} (\bibinfo {year} {2004})}\BibitemShut {NoStop}%
  \bibitem [{Note1()}]{SM}%
  \BibitemOpen
  \bibinfo {note} {See Supplementary Material for details on MF qubit basis, Schrieffer-Wolff transformation, $J_{\kappa\lambda}$, canonical form of $U_\textrm{CNOT}$, $H^{(12)}_{MF}$, and parity readout of the TSCs.}\BibitemShut {Stop}%
\bibitem [{\citenamefont {Schrieffer}\ and\ \citenamefont
  {Wolff}(1966)}]{schriefferPR66}%
  \BibitemOpen
  \bibfield  {author} {\bibinfo {author} {\bibfnamefont {J.}~\bibnamefont
  {Schrieffer}}\ and\ \bibinfo {author} {\bibfnamefont {P.}~\bibnamefont
  {Wolff}},\ }\href@noop {} {\bibfield  {journal} {\bibinfo  {journal} {Phys.
  Rev.}\ }\textbf {\bibinfo {volume} {149}},\ \bibinfo {pages} {491} (\bibinfo
  {year} {1966})}\BibitemShut {NoStop}%
 \bibitem [{\citenamefont {Lee}\ \emph {et~al.}(2013)\citenamefont {Lee},
  \citenamefont {Lim},\ and\ \citenamefont {L{\'o}pez}}]{leePRB13}%
  \BibitemOpen
  \bibfield  {author} {\bibinfo {author} {\bibfnamefont {M.}~\bibnamefont
  {Lee}}, \bibinfo {author} {\bibfnamefont {J.~S.}\ \bibnamefont {Lim}}, \ and\
  \bibinfo {author} {\bibfnamefont {R.}~\bibnamefont {L{\'o}pez}},\ }\href@noop
  {} {\bibfield  {journal} {\bibinfo  {journal} {Phys. Rev. B}\ }\textbf
  {\bibinfo {volume} {87}},\ \bibinfo {pages} {241402} (\bibinfo {year}
  {2013})}\BibitemShut {NoStop}%
 \bibitem [{\citenamefont {Vernek}\ \emph {et~al.}(2014)\citenamefont {Vernek},
  \citenamefont {Penteado}, \citenamefont {Seridonio},\ and\ \citenamefont
  {Egues}}]{vernekPRB14}%
  \BibitemOpen
  \bibfield  {author} {\bibinfo {author} {\bibfnamefont {E.}~\bibnamefont
  {Vernek}}, \bibinfo {author} {\bibfnamefont {P.}~\bibnamefont {Penteado}},
  \bibinfo {author} {\bibfnamefont {A.}~\bibnamefont {Seridonio}}, \ and\
  \bibinfo {author} {\bibfnamefont {J.}~\bibnamefont {Egues}},\ }\href@noop {}
  {\bibfield  {journal} {\bibinfo  {journal} {Phys. Rev. B}\ }\textbf
  {\bibinfo {volume} {89}},\ \bibinfo {pages} {165314} (\bibinfo {year}
  {2014})}\BibitemShut {NoStop}%
\bibitem [{\citenamefont {Nielsen}\ and\ \citenamefont
  {Chuang}(2010)}]{nielsenBK10}%
  \BibitemOpen
  \bibfield  {author} {\bibinfo {author} {\bibfnamefont {M.~A.}\ \bibnamefont
  {Nielsen}}\ and\ \bibinfo {author} {\bibfnamefont {I.~L.}\ \bibnamefont
  {Chuang}},\ }\href@noop {} {\emph {\bibinfo {title} {Quantum Computation and
  Quantum Information}}}\ (\bibinfo  {publisher} {Cambridge University Press},\
  \bibinfo {year} {2010})\BibitemShut {NoStop}%
 \bibitem [{\citenamefont {Trifunovic}\ \emph {et~al.}(2012)\citenamefont
  {Trifunovic}, \citenamefont {Dial}, \citenamefont {Trif}, \citenamefont
  {Wootton}, \citenamefont {Abebe}, \citenamefont {Yacoby},\ and\ \citenamefont
  {Loss}}]{trifunovicPRX12}%
  \BibitemOpen
  \bibfield  {author} {\bibinfo {author} {\bibfnamefont {L.}~\bibnamefont
  {Trifunovic}}, \bibinfo {author} {\bibfnamefont {O.}~\bibnamefont {Dial}},
  \bibinfo {author} {\bibfnamefont {M.}~\bibnamefont {Trif}}, \bibinfo {author}
  {\bibfnamefont {J.~R.}\ \bibnamefont {Wootton}}, \bibinfo {author}
  {\bibfnamefont {R.}~\bibnamefont {Abebe}}, \bibinfo {author} {\bibfnamefont
  {A.}~\bibnamefont {Yacoby}}, \ and\ \bibinfo {author} {\bibfnamefont
  {D.}~\bibnamefont {Loss}},\ }\href@noop {} {\bibfield  {journal} {\bibinfo
  {journal} {Phys. Rev. X}\ }\textbf {\bibinfo {volume} {2}},\ \bibinfo {pages}
  {011006} (\bibinfo {year} {2012})}\BibitemShut {NoStop}%
\bibitem [{\citenamefont {Alicea}\ \emph {et~al.}(2011)\citenamefont {Alicea},
  \citenamefont {Oreg}, \citenamefont {Refael}, \citenamefont {von Oppen},\
  and\ \citenamefont {Fisher}}]{aliceaNATP11}%
  \BibitemOpen
  \bibfield  {author} {\bibinfo {author} {\bibfnamefont {J.}~\bibnamefont
  {Alicea}}, \bibinfo {author} {\bibfnamefont {Y.}~\bibnamefont {Oreg}},
  \bibinfo {author} {\bibfnamefont {G.}~\bibnamefont {Refael}}, \bibinfo
  {author} {\bibfnamefont {F.}~\bibnamefont {von Oppen}}, \ and\ \bibinfo
  {author} {\bibfnamefont {M.~P.~A.}\ \bibnamefont {Fisher}},\ }\href
  {http://dx.doi.org/10.1038/nphys1915} {\bibfield  {journal} {\bibinfo
  {journal} {Nat. Phys.}\ }\textbf {\bibinfo {volume} {7}},\ \bibinfo {pages}
  {412} (\bibinfo {year} {2011})}\BibitemShut {NoStop}%
  \bibitem [{\citenamefont {Raussendorf}\ and\ \citenamefont
  {Harrington}(2007)}]{raussendorfPRL07}%
  \BibitemOpen
  \bibfield  {author} {\bibinfo {author} {\bibfnamefont {R.}~\bibnamefont
  {Raussendorf}}\ and\ \bibinfo {author} {\bibfnamefont {J.}~\bibnamefont
  {Harrington}},\ }\href@noop {} {\bibfield  {journal} {\bibinfo  {journal}
  {Phys. Rev. Lett.}\ }\textbf {\bibinfo {volume} {98}},\ \bibinfo {pages}
  {190504} (\bibinfo {year} {2007})}\BibitemShut {NoStop}%
\bibitem [{\citenamefont {Wang}\ \emph {et~al.}(2011)\citenamefont {Wang},
  \citenamefont {Fowler},\ and\ \citenamefont {Hollenberg}}]{wangPRA11}%
  \BibitemOpen
  \bibfield  {author} {\bibinfo {author} {\bibfnamefont {D.~S.}\ \bibnamefont
  {Wang}}, \bibinfo {author} {\bibfnamefont {A.~G.}\ \bibnamefont {Fowler}}, \
  and\ \bibinfo {author} {\bibfnamefont {L.~C.}\ \bibnamefont {Hollenberg}},\
  }\href@noop {} {\bibfield  {journal} {\bibinfo  {journal} {Phys. Rev. A}\
  }\textbf {\bibinfo {volume} {83}},\ \bibinfo {pages} {020302} (\bibinfo
  {year} {2011})}\BibitemShut {NoStop}%
\bibitem [{\citenamefont {Lee}\ \emph {et~al.}(2012)\citenamefont {Lee},
  \citenamefont {Jiang}, \citenamefont {Aguado}, \citenamefont {Katsaros},
  \citenamefont {Lieber},\ and\ \citenamefont {De~Franceschi}}]{leePRL12}%
  \BibitemOpen
  \bibfield  {author} {\bibinfo {author} {\bibfnamefont {E.~J.}\ \bibnamefont
  {Lee}}, \bibinfo {author} {\bibfnamefont {X.}~\bibnamefont {Jiang}}, \bibinfo
  {author} {\bibfnamefont {R.}~\bibnamefont {Aguado}}, \bibinfo {author}
  {\bibfnamefont {G.}~\bibnamefont {Katsaros}}, \bibinfo {author}
  {\bibfnamefont {C.~M.}\ \bibnamefont {Lieber}}, \ and\ \bibinfo {author}
  {\bibfnamefont {S.}~\bibnamefont {De~Franceschi}},\ }\href@noop {} {\bibfield
   {journal} {\bibinfo  {journal} {Phys. Rev. Lett.}\ }\textbf {\bibinfo
  {volume} {109}},\ \bibinfo {pages} {186802} (\bibinfo {year}
  {2012})}\BibitemShut {NoStop}%
\bibitem [{\citenamefont {Fasth}\ \emph {et~al.}(2007)\citenamefont {Fasth},
  \citenamefont {Fuhrer}, \citenamefont {Samuelson}, \citenamefont {Golovach},\
  and\ \citenamefont {Loss}}]{fasthPRL07}%
  \BibitemOpen
  \bibfield  {author} {\bibinfo {author} {\bibfnamefont {C.}~\bibnamefont
  {Fasth}}, \bibinfo {author} {\bibfnamefont {A.}~\bibnamefont {Fuhrer}},
  \bibinfo {author} {\bibfnamefont {L.}~\bibnamefont {Samuelson}}, \bibinfo
  {author} {\bibfnamefont {V.~N.}\ \bibnamefont {Golovach}}, \ and\ \bibinfo
  {author} {\bibfnamefont {D.}~\bibnamefont {Loss}},\ }\href@noop {} {\bibfield
   {journal} {\bibinfo  {journal} {Phys. Rev. Lett.}\ }\textbf {\bibinfo
  {volume} {98}},\ \bibinfo {pages} {266801} (\bibinfo {year}
  {2007})}\BibitemShut {NoStop}%
\bibitem [{\citenamefont {Nadj-Perge}\ \emph {et~al.}(2010)\citenamefont
  {Nadj-Perge}, \citenamefont {Frolov}, \citenamefont {Bakkers},\ and\
  \citenamefont {Kouwenhoven}}]{nadj-pergeNAT10}%
  \BibitemOpen
  \bibfield  {author} {\bibinfo {author} {\bibfnamefont {S.}~\bibnamefont
  {Nadj-Perge}}, \bibinfo {author} {\bibfnamefont {S.}~\bibnamefont {Frolov}},
  \bibinfo {author} {\bibfnamefont {E.}~\bibnamefont {Bakkers}}, \ and\
  \bibinfo {author} {\bibfnamefont {L.~P.}\ \bibnamefont {Kouwenhoven}},\
  }\href@noop {} {\bibfield  {journal} {\bibinfo  {journal} {Nature}\ }\textbf
  {\bibinfo {volume} {468}},\ \bibinfo {pages} {1084} (\bibinfo {year}
  {2010})}\BibitemShut {NoStop}%
\bibitem [{\citenamefont {Nadj-Perge}\ \emph {et~al.}(2012)\citenamefont
  {Nadj-Perge}, \citenamefont {Pribiag}, \citenamefont {Van~den Berg},
  \citenamefont {Zuo}, \citenamefont {Plissard}, \citenamefont {Bakkers},
  \citenamefont {Frolov},\ and\ \citenamefont {Kouwenhoven}}]{nadj-pergePRL12}%
  \BibitemOpen
  \bibfield  {author} {\bibinfo {author} {\bibfnamefont {S.}~\bibnamefont
  {Nadj-Perge}}, \bibinfo {author} {\bibfnamefont {V.}~\bibnamefont {Pribiag}},
  \bibinfo {author} {\bibfnamefont {J.}~\bibnamefont {Van~den Berg}}, \bibinfo
  {author} {\bibfnamefont {K.}~\bibnamefont {Zuo}}, \bibinfo {author}
  {\bibfnamefont {S.}~\bibnamefont {Plissard}}, \bibinfo {author}
  {\bibfnamefont {E.}~\bibnamefont {Bakkers}}, \bibinfo {author} {\bibfnamefont
  {S.}~\bibnamefont {Frolov}}, \ and\ \bibinfo {author} {\bibfnamefont
  {L.}~\bibnamefont {Kouwenhoven}},\ }\href@noop {} {\bibfield  {journal}
  {\bibinfo  {journal} {Phys. Rev. Lett.}\ }\textbf {\bibinfo {volume} {108}},\
  \bibinfo {pages} {166801} (\bibinfo {year} {2012})}\BibitemShut {NoStop}%
\bibitem [{\citenamefont {De~Franceschi}\ \emph {et~al.}(2010)\citenamefont
  {De~Franceschi}, \citenamefont {Kouwenhoven}, \citenamefont {Schonenberger},\
  and\ \citenamefont {Wernsdorfer}}]{defranceschiNATN10}%
  \BibitemOpen
  \bibfield  {author} {\bibinfo {author} {\bibfnamefont {S.}~\bibnamefont
  {De~Franceschi}}, \bibinfo {author} {\bibfnamefont {L.}~\bibnamefont
  {Kouwenhoven}}, \bibinfo {author} {\bibfnamefont {C.}~\bibnamefont
  {Schonenberger}}, \ and\ \bibinfo {author} {\bibfnamefont {W.}~\bibnamefont
  {Wernsdorfer}},\ }\href {http://dx.doi.org/10.1038/nnano.2010.173} {\bibfield
   {journal} {\bibinfo  {journal} {Nat. Nano.}\ }\textbf {\bibinfo {volume}
  {5}},\ \bibinfo {pages} {703} (\bibinfo {year} {2010})}\BibitemShut {NoStop}%
\bibitem [{\citenamefont {Szombati}\ \emph {et~al.}(2015)\citenamefont
  {Szombati}, \citenamefont {Nadj-Perge}, \citenamefont {Car}, \citenamefont
  {Plissard}, \citenamefont {Bakkers},\ and\ \citenamefont
  {Kouwenhoven}}]{szombatiCM15}%
  \BibitemOpen
  \bibfield  {author} {\bibinfo {author} {\bibfnamefont {D.}~\bibnamefont
  {Szombati}}, \bibinfo {author} {\bibfnamefont {S.}~\bibnamefont
  {Nadj-Perge}}, \bibinfo {author} {\bibfnamefont {D.}~\bibnamefont {Car}},
  \bibinfo {author} {\bibfnamefont {S.}~\bibnamefont {Plissard}}, \bibinfo
  {author} {\bibfnamefont {E.}~\bibnamefont {Bakkers}}, \ and\ \bibinfo
  {author} {\bibfnamefont {L.}~\bibnamefont {Kouwenhoven}},\ }\href@noop {}
  {\bibfield  {journal} {\bibinfo  {journal} {arXiv preprint arXiv:1512.01234}\
  } (\bibinfo {year} {2015})}\BibitemShut {NoStop}%
\end{thebibliography}
\end{document}